\documentclass[reprint, amsmath, amssymb, aps, prl
, longbibliography
, nofootinbib
]{revtex4-1}

\usepackage{graphicx}
\usepackage[dvipsnames]{xcolor}
\newcommand{\authormain}{Farshid Jafarpour}
\newcommand{\titlemain}{Cell-size-regulation-induced oscillations in population growth rate}
\usepackage[bookmarks={false}, pdfauthor={\authormain}, pdftitle={\titlemain}]{hyperref}
\hypersetup{colorlinks=true, linkcolor=BrickRed, citecolor=Violet, filecolor=OliveGreen, urlcolor=RoyalBlue, filebordercolor={.8 .8 1}, urlbordercolor={.8 .8 0}}

\newcommand\Ref[1]{Ref.~\cite{#1}}
\newcommand\eq[1]{Eq.~(\ref{eq:#1})}
\newcommand\Eq[1]{Equation~(\ref{eq:#1})}
\newcommand\fig[1]{Fig.~\ref{fig:#1}}
\newcommand\Fig[1]{Figure~\ref{fig:#1}}

\newcommand\mean[1]{\left\langle#1\right\rangle}

\begin{document}
\title{Cell size regulation induces sustained oscillations in the population growth rate}

\author{Farshid Jafarpour}
\affiliation{University of Pennsylvania Department of Physics \& Astronomy,
209 South 33rd Street, Philadelphia, PA 19104-6396}
\date{\today}

\begin{abstract}
We study the effect of correlations in generation times on the dynamics of population growth of microorganisms. We show that any non-zero correlation that is due to cell-size regulation, no matter how small, induces long-term oscillations in the population growth rate. The population only reaches its steady state when we include the often-neglected variability in the growth rates of individual cells. We discover that the relaxation time scale of the population to its steady state is determined by the distribution of single-cell growth rates and is surprisingly independent of details of the division process such as the noise in the timing of division and the mechanism of cell-size regulation. We validate the predictions of our model using existing experimental data and propose an experimental method to measure single-cell growth variability by observing how long it takes for the population to reach its steady state or balanced growth.  
\end{abstract}

\maketitle
Most of us have first cousins that are more or less our age, but the ages of our more distant cousins are more broadly distributed. The difference arises due to the larger number of generations since our last common ancestor with our more distant cousins. The noise in the generation times adds up over generations, giving rise to wider distributions of ages. The number of generations it takes for the descendants of an individual to sufficiently mix in age to be statistically indistinguishable from the rest of the population is inversely related to the variability in the generation times~\cite{jafarpour2018bridging}.  

\begin{figure}[b]
\vspace{-1.0em}
	\includegraphics[width=\columnwidth]{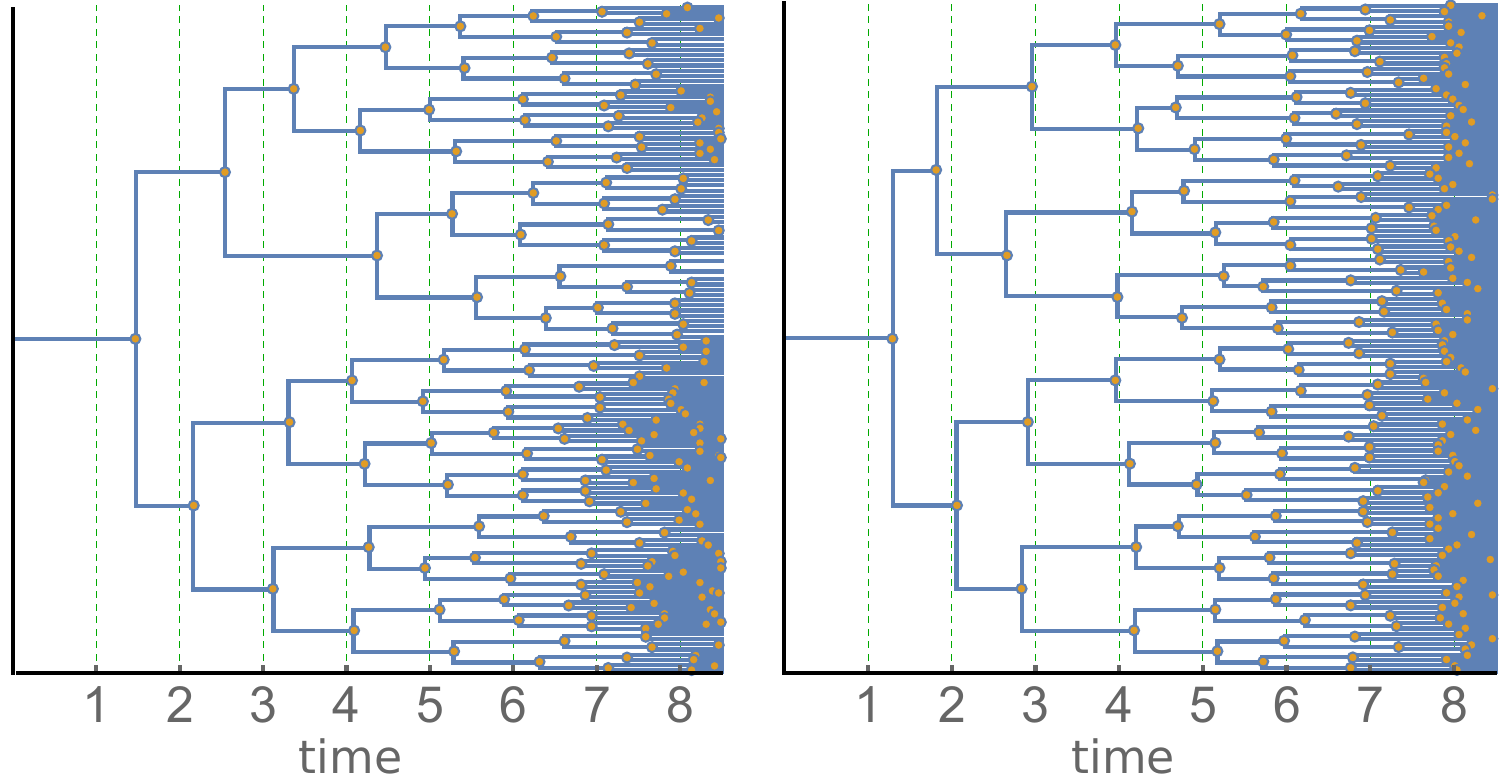}
	\caption{Lineage tree of populations starting from a single cell. (left) In the absence of cell size control, the division times (circular markers) become less synchronized over time due to the accumulation of noise in their generation times. (right) The division times of cells with cell size control stay synchronized due to correlations in the generation times of mother and daughter cells.}\vspace{-0.5em}
	\label{fig:tree}
\end{figure}

Here, we show that this problem is very different in the context of single cellular organisms due to the interaction between cell size and generation time. Many single cellular organisms grow exponentially in size before division~\cite{wang2010robust, campos2014constant,iyer2014scaling, iyer2014universality, taheri2015cell,pirjol2017phenomenology, di2007effects, eun2018archaeal}. If a cell grows for a longer time than expected before it divides, its daughter cells will be larger at birth and have to compensate for their sizes by dividing slightly earlier than expected. Otherwise, the noise in the generation times would accumulate over generations in the size of the cells, leading to extremely large cells~\cite{amir2014cell}. This compensation for the error in the generation times not only suppresses the accumulation of noise in cell sizes, but also prevents the accumulation of noise in the distribution of ages over generations and keeps the division times synchronized (see \fig{tree}). Given this observation, it is natural to ask what sets the time scale for a population of microorganisms to desynchronize and reach its steady state.

In this Letter, we study the dynamics of population growth of microorganisms starting from a single cell. We show that the correlations induced by the cell-size control mechanism, no matter how small, significantly delay the relaxation of the population to its steady state. We observe transient oscillations in the growth rate of the number of cells in the population. These oscillations are sustained by the mother-daughter correlations and decay due to the competing effect of small variations in the single cell growth rates. We discover that the single cell growth rate distribution completely determines the timescale for the relaxation of the population to its steady state as well as the steady state population growth rate irrespective of the details of cell division process and cell-size control mechanism. 

The distribution of single-cell growth rates is a major evolutionary trait contributing to the fitness of an organism~\cite{de2016growth, nozoe2017inferring, robert2018mutation}. It has been recently shown that the steady-state growth rate of a population can be found from the distribution of single-cell growth rates~\cite{lin2017effects,lin2018population}. Since the population growth rate is easier to measure than the single cell growth rate distribution, it would be desirable to go in the reverse direction. We provide a relationship between the decay rate of the oscillations in the growth rate of the population and the distribution of single-cell growth rates. This relationship can be used in combination with steady-state results to estimate the growth rate distribution by observing the growth of a population as it relaxes to its steady state. We validate this prediction using the existing single cells data from the \lq\lq mother machine" experiment from \Ref{wang2010robust}.

\textit{Theoretical Model}: We use a model introduced in \Ref{amir2014cell}, where cells grow exponentially in size with growth rate $\kappa$. Each cell with birth size $v_b$ attempts to divide after its size reaches a target size $v_d = f(v_b)$. We assume a time additive\footnote
	{Realistically, this noise would be multiplicative. For small noise, the multiplicative factor can be evaluated at the expected value leading to an additive noise. One could argue that it is more natural to use a size-additive noise since cells control the timing of their division based on their size. Simulations results (not shown here) indicate that the dynamics is not affected if we use size-additive noise instead.} 
noise $\xi$ in the division process with zero mean and variance $\sigma_\xi^2$ such that the generation time $\tau$ is given by
\begin{equation}
	\tau = \frac1\kappa\ln\left(\frac{v_d}{v_b}\right)+\xi.
\end{equation}
The function $f(v_b)$ determines the cell-size control mechanism. In the presence of cell-size control, the sequence of initial sizes, $v_b^{n+1}=f(v_b^n)/2$ has a fixed point $\Delta$, and the distribution of initial cell sizes is sharply peaked around $\Delta$. Therefore, all reasonable functions $f$ that are equivalent to linear order near $\Delta$ describe approximately the same dynamics. The one parameter family of functions $f(v_b) = 2 \Delta^\alpha v_b^{1-\alpha}$, with $0\leq \alpha \leq 1$ qualitatively captures the full range of behavior for this model and interpolates\footnote{If we linearize $f$ around its fixed point $\Delta$, we obtain $f(v_b) \approx 2\alpha\Delta+2(1-\alpha)v_b$ which is a linear interpolation
	between the timer and the sizer model. This convenient nonlinear form simplifies analytical calculations while leaving the simulation results virtually 		unchanged. This is due to the narrow distribution of $v_b$ around the fixed point $\Delta$.}
between two extremes~\cite{amir2014cell, ho2018modeling}. The case $\alpha=0$, known as the timer model, has no cell size control where cells attempt to divide after a period of time independent of their size. Successive generation times in this case are uncorrelated with variance $\sigma_\xi^2$ while the variance of the cell size distribution is known to diverge at long time~\cite{amir2014cell}. The case $\alpha=1$ is known as the sizer model where cells attempt to divide when they reach the size $2\Delta$ independent of their history~\cite{diekmann1983growth}. The variance in the generation times in this case is given by $2\sigma_\xi^2$. Experimental data support a value of $\alpha$ closer to $1/2$ for many organisms, where cells attempt to divide when they approximately add a constant size $\Delta$ to their original size~\cite{amir2014cell,campos2014constant,taheri2015cell,jun2015cell, sauls2016adder,logsdon2017parallel,jun2018fundamental,eun2018archaeal,banerjee2017biphasic,si2018mechanistic}. For $\alpha>0$, the generation times of mother and daughter cells are correlated with Pearson correlation coefficient $C_{MD} = -\alpha/2$~\cite{lin2017effects}, and variance of the generation time is given by $2\sigma_\xi^2/(2-\alpha)$.

\begin{figure}[t]
	\includegraphics[width=\columnwidth]{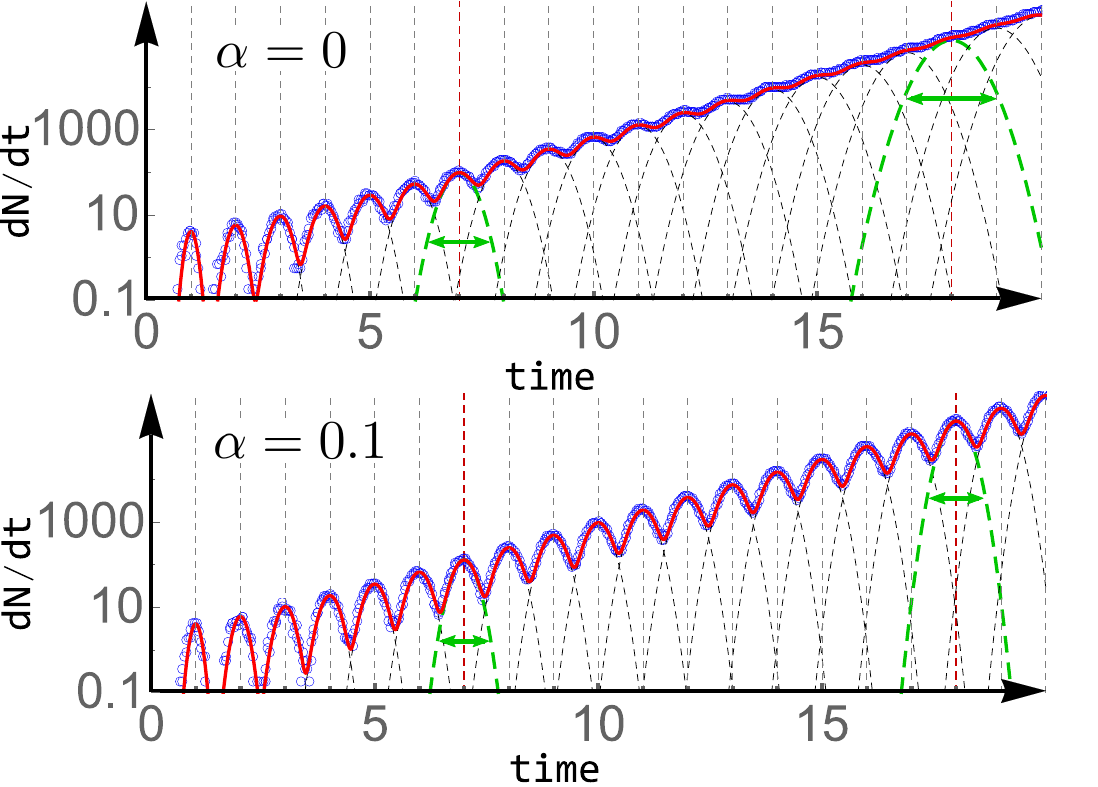}
	\caption{Log-scale plot of the expected value of the rate of change of the number of cells in a population starting with a single cell, calculated analytically (red solid curve) and compared with simulation (blue circles). The rate of change of the number of cells can be written as the sum of the division rates (parabolic dashed lines) of all generations (see \eq{delta_t}). (Top) In the absence of cell size control, $\alpha=0$, the distribution of division times of higher generations get wider and start to overlap, damping out the oscillations in the growth rate. (Bottom) In the presence of even a small cell size control, $\alpha=0.1$, the distribution of successive division times quickly approach a steady state distribution with a finite variance (see \eq{var}) leading to the persistence of oscillations in the growth of the population. The distribution of timing of the $7$th and $18$th generations are highlighted in both cases for comparison.}
	\label{fig:dNdt}
\end{figure}

\textit{Transient Oscillations in the Absence of Cell-Size Control}: In the absence of cell-size control ($\alpha=0$), the correlation between the division times of mother and daughter cells is zero and the dynamics of the age distribution (the distribution of time since last division) decouples from the size distribution. This case and its generalization to asymmetric division are studied in \Ref{jafarpour2018bridging}. For a population starting from a single cell, the timing of the $n$th division, $t_n$, is given by
\begin{equation}
	t_n = \sum_{i=1}^n \tau_i = n \bar\tau + \sum_{i=1}^n \xi_i,
\end{equation}
where $\bar \tau=\ln(2)/\kappa$ is the cell-size doubling time. The distribution of $t_n$ has the mean $n \bar \tau$ and the variance $n\sigma_\xi^2$. Because the distribution of generation times is often peaked sharply around its mean, the growth in the number of cells happens in bursts of synchronized divisions around the times $\mean{t_n} = n\bar\tau$, leading to oscillating behavior in the growth rate and the traveling waves in the age-distribution~\cite{jafarpour2018bridging}. As the noise in the division process accumulates in successive division events, the division bursts widen and start to overlap, and as a result, the population starts to desynchronize and the age-distribution relaxes to its steady state (see the top panel of \fig{dNdt}).

\textit{Sustained Oscillations in the Presence of Cell-Size Control}: Now we consider the general case of non-zero $\alpha$ and show that the accumulation of noise is suppressed due to the negative correlation induced by the cell size control\footnote{Negative correlations between the generation times 
	of mother 	and daughter cells that are not due to cell size control are
	not sufficient to sustain these oscillations. This can be seen by considering
	a model in which a negative correlation is artificially imposed on 
	the generation times by setting $\tau_{i+1} = \bar{\tau} 
	- \alpha (\tau_i - \bar{\tau}) + \xi_i$ independent of the size of the 
	cell. In this model, for all values of $\alpha>0$, the accumulation of
	the noise is only partially suppressed and the oscillations
	decay at long time.} 
leading to the persistence of the oscillations in the growth rate and traveling waves in the age-structure of the population at long time. In this case, the timing of the $n$th generation division can be written as $t_n = n\bar \tau + \delta t_n$ where $\delta t_n$s are random variables with probability density $g_n(\delta t)$. We have derived the following recursive relationship for $g_n$ (See the Supplemental Material (SM)~\cite{SM} for the derivation)
\begin{equation}\label{eq:delta_t}
	g_n(\delta t) = \int g_{n-1}(\delta t-\delta \tau) f_\xi\big((1-\alpha)\delta\tau+\alpha \delta t\big) d\delta\tau,
\end{equation}
where $f_\xi$ is the probability density function of $\xi$. Now, the expected value of the rate of change in the total number of cells in the population can be written as a sum over the division rates (number of cells produced in each generation multiplied by the division time distribution) of all the generations 
\begin{equation}
	\frac{dN}{dt} = \sum_{n=1}^\infty 2^n g_n(t-n\bar\tau).
\end{equation}
The width of the distribution of $\delta t$ is the key to understanding the sustained oscillations. For $\alpha=0$, the integral in \eq{delta_t} becomes a convolution leading to the accumulation of the noise at each generation. For $\alpha=1$, the distribution of $\delta t$ is given by $f_\xi$ independent of the generation number $n$. The cells in this case do not desynchronize at later generations and the change in the population is characterized by periodic bursts of divisions at regular intervals. For $0<\alpha\leq1$, the variance of the division time at the $n$th generation is found using \eq{delta_t} to be
\begin{equation}\label{eq:var}
	\text{var}(t_n) = \sigma_\xi^2 \frac{1-(1-\alpha)^{2n}}{\alpha (2-\alpha)}.
\end{equation}
For $\alpha>0$, the successive division time distributions approach a limiting distribution with the finite variance $\sigma_\xi^2/\alpha(2-\alpha)$, leading to periodic bursts of divisions and oscillatory behavior. Here, we have made no assumption about the distribution of the noise $\xi$ in the timing of the division process except that it has a finite variance. For a more concrete example, let us consider a Gaussian form for $f_\xi$\footnote{We assume the variance is small enough so that $\tau$ does not become negative}. In this case, using \eq{delta_t}, we are able to show that the $g_n$s are also Gaussian distributed with the variance given in \eq{var}. \Fig{dNdt} shows the comparison of these analytical results with the numerical simulations for two cases of no cell size control, $\alpha=0$, and a small cell size control, $\alpha=0.1$. 
\begin{figure}[t]
	\includegraphics[width=\columnwidth]{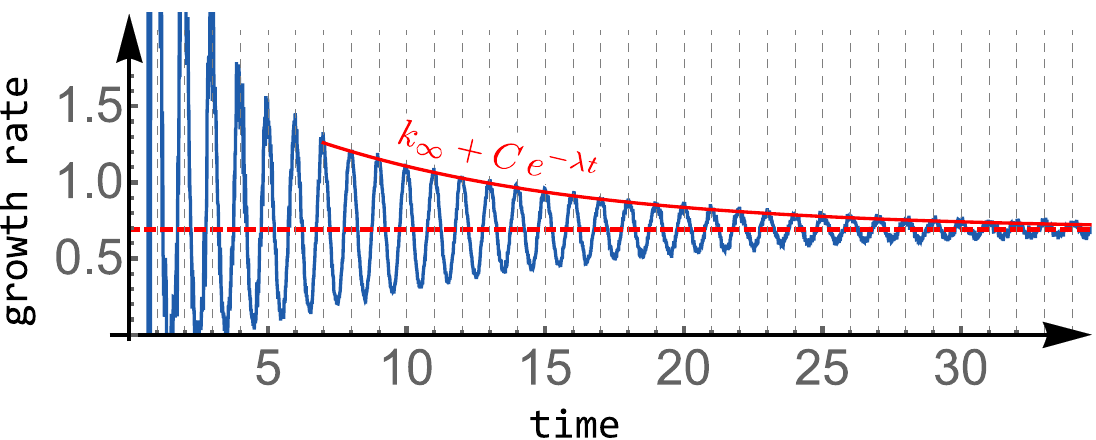}
	\caption{Oscillations in the population growth rate decay exponentially due to the stochasticity in growth rates of individual cells. The solid red line is the exponential fit used in \fig{decay_rate}, the horizontal dashed red line is the steady-state value of the population growth rate, and vertical dashed lines are the expected values of successive division times where the population growth rate peaks. Simulation parameters: $\alpha = 0.5$, $\sigma_\xi = 0.1$, $\bar\kappa=\ln(2)$, and $\sigma_\kappa = 0.07\bar \kappa$.}
	\label{fig:decaying}
\end{figure}

As shown in \Ref{jafarpour2018bridging} the synchronized bursts of division and oscillations in the population growth rate are associated with traveling waves in the age distribution. Since sizes of cells grow exponentially with their age, the traveling waves in the age distribution induce similar waves in the size distribution of the population which can be used as an alternative method to detect these oscillations.

\textit{Stochastic Growth Rate}: In practice, a population of uncoupled cells cannot maintain synchronized division for infinite time and the oscillations in the growth rates have to decay as the population relaxes to its steady-state age distribution. In order to capture this relaxation and estimate its time scale, we need to include multiple sources of noise in our model. There are at least two other sources of stochasticity in the growth and division of cells: (1) small variability in the growth rate of the individual cells from one generation to another and (2) random asymmetry in the division plane of otherwise symmetrically dividing cells. In many symmetrically dividing organisms, the coefficient of variation (CV, the ratio of the standard deviation to the mean) of the single-cell growth rate, $\kappa$, is significantly larger than that of the division ratio (DR, the ratio of the size of the daughter cell to that of its mother cell). For example in \textit{E. coli}, the CV of DR is between $0.02$ and $0.06$~\cite{guberman2008psicic,eun2018archaeal,campos2014constant} while the CV of single cell growth rate is reported to be between $0.06$ and $0.20$ depending on the growth condition~\cite{lin2017effects,cermak2016high,wallden2016synchronization,kennard2016individuality}.  Here we consider organisms in which the stochasticity in the DR can be neglected. The case with stochastic DR is studied in SM~\cite{SM}. Since $\kappa$ has a narrow symmetric distribution around its mean $\bar \kappa$~\cite{kennard2016individuality}, its distribution can be estimated as a Gaussian with some variance $\sigma_\kappa^2$. Furthermore, unlike the correlation in the generation times, the correlation between the growth rate of mother and daughter cells can be negligible depending on the organism and the growth condition~\cite{wang2010robust,grilli2018empirical} and are ignored in this model.

\Fig{decaying} shows the population growth rate, $k \equiv d\ln(N)/dt$, in a simulation of the model described above, where now the growth rate of each cell is independently chosen from a Gaussian distribution with the mean $\bar \kappa=\ln(2)$ and CV of $0.07$ (time is measured in the unit of $\bar \tau= \ln(2)/\bar \kappa$). We observe that oscillations in the population growth rate decay exponentially at long time until the growth rate approaches a steady state value. This value is given by the unique $k$ satisfying the equation
\begin{equation}\label{eq:k-exact}
	\mean{ \left(\frac12\right)^{k/\kappa}}_\kappa \equiv \int_0^\infty \rho(\kappa)\, 2^{-k/\kappa} d\kappa = \frac12
\end{equation}
where $\rho(\kappa)$ is the distribution of single-cell growth rates. This relationship is obtained from Powell's relationship for population growth rate in the absence of correlations~\cite{powell1956growth}, by replacing the distribution of generation times with the distribution of cell size doubling times. This can be done because the noise in the timing of division does not play a role in the population growth rate in the presence of cell size control~\cite{lin2017effects} (see SM~\cite{SM} for comparison to simulation). In \eq{k-exact}, $\rho$ is the distribution along a lineage (or equivalently over the entire population tree) which is distinct from the instantaneous population distribution~\cite{lin2017effects,rochman2018ergodicity,susman2018individuality}. Since the slow-growing cells have longer generation times, they are overrepresented in the population at any given time, and therefore, the population growth rate is always slightly smaller than $\bar \kappa$ \footnote{This argument fails if the growth rates of mother and daughter cells are highly correlated, in which case the fast-growing cells reproduce faster and can in principle compensate for their under-representation in the population depending on the strength of the correlation. See \Ref{lin2018population} for a more detailed discussion.}. For a narrow distribution, the population growth rate can be approximated in terms of $\bar \kappa$ and $\sigma_\kappa$~\cite{lin2018population}
\begin{equation}\label{eq:k-approx}
	k \approx \bar \kappa - \left(1-\frac{\ln(2)}{2}\right)\frac{\sigma_\kappa^2}{\bar \kappa}.
\end{equation}

\begin{figure}[t]
	\includegraphics[width=\columnwidth]{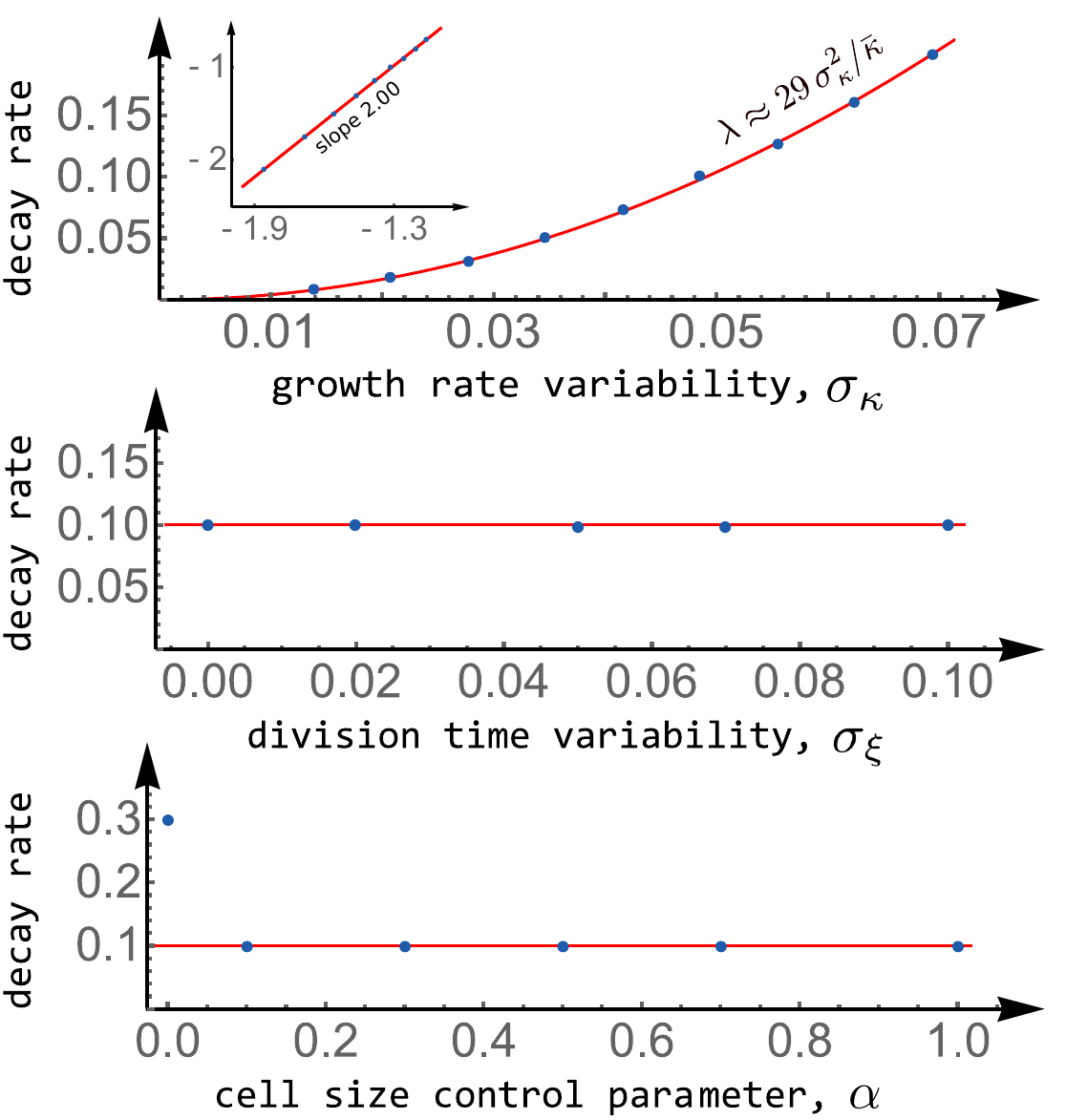}
	\caption{Simulation results for the decay rate of oscillations in the population growth rate shown as functions of $\sigma_\kappa$, $\sigma_\xi$, and $\alpha$: (top) decay rate increases linearly with the variance of single-cell growth rate distribution, $\sigma_\kappa^2$ (the red solid line is a parabolic fit; inset is the log-log plot with linear fit); (middle) the noise in the division process has no effect on the damping of the oscillations in the population growth rate; (bottom) the mechanism for cell-size control does not affect the decay rate either as long as there is a nonzero cell-size control, $\alpha\neq 0$. Time is measured in units of $\bar\tau$. Simulation parameters: (top) $\alpha = 0.5$ and $\sigma_\xi = 0.1\bar\tau$, (middle) $\alpha = 0.5$ and $\sigma_\kappa = 0.07\bar \kappa$, and (bottom) $\sigma_\xi = 0.1\bar\tau$ and $\sigma_\kappa = 0.07\bar \kappa$.}
	\label{fig:decay_rate}
\end{figure}

We have a total of five independent variables in our model: $\alpha$, $\sigma_\xi$, $\sigma_\kappa$, $\Delta$, and $\bar\tau = \ln(2)/\bar \kappa$. Time and size can be measured in units of $\bar\tau$ and $\Delta$, respectively. \Fig{decay_rate} shows the dependence of the rate of decay of the oscillations of the population growth rate on all of the remaining model parameters $\alpha$, $\sigma_\xi$, and $\sigma_\kappa$. Surprisingly, this decay rate is completely independent of the mechanism of cell size control, $\alpha$ (with the exception of the single point $\alpha=0$), and is also independent of the noise in the timing of the division process, $\sigma_\xi$.  It is proportional to the variance of the single-cell growth rate distribution, $\sigma_\kappa^2$. An experimental measurement of this decay rate in the oscillations of population growth rate using cell counting techniques can provide the width of the single-cell growth rate distribution. 

\textit{Conclusion}: For nearly a century, microbiologists have been concerned with the relationship between statistical observables of single cells and the properties of their populations~\cite{m1925applications,powell1956growth,von1959kinetics,painter1968mathematics,painter1968mathematics, perthame2006transport,rubin2016mathematical,jun2018fundamental}. Recent advances in single cell tracking technology have lead to a surge of renewed interest in this field~\cite{wakamoto2005single,siegal2008tightly,wang2010robust,sliusarenko2011high,young2012measuring,salman2012universal, lambert2015quantifying}. On one hand, the details of the mechanism of cell size control that allows populations to maintain a narrow distribution of cell sizes~\cite{schaechter1962growth,koch1962model,anderson1969cell,painter1968mathematics,fantes1975regulation,jorgensen2004cells,chien2012cell,turner2012cell,lloyd2013regulation} has become the topic of an intense debate over the past few years~\cite{osella2014concerted,robert2014division,osella2014concerted,amir2014cell,campos2014constant,taheri2015cell,marantan2016stochastic,wallden2016synchronization, grilli2017relevant,ho2018modeling,cadart2018size}. On the other hand, the relationship between the stochasticity in the generation times of microorganisms and the growth of their populations has gained recent attention~\cite{hashimoto2016noise,cerulus2016noise,kennard2016individuality,rochman2018ergodicity,susman2018individuality,gavagnin2018invasion, jafarpour2018bridging}. There are two distinct sources of stochasticity in generation times: the noise in the cellular growth and the noise in the division process. We claim that only the former plays a role in the growth and relaxation rates of the population, while cell size control is precisely the process of canceling out the latter over the course of a few generations. As a result, both the time it takes for a population to reach its steady state and the steady-state population growth rate is only affected by the portion of noise in the generation times that is due to the variability in the single cell growth rates and not the stochasticity in the timing of cell division.

The distribution of single-cell growth rates is a major evolutionary trait contributing to the fitness of an organism. It fully determines the steady state population growth rate irrespective of the details of cell division process including both the noise in the timing of division and the details of cell size control mechanism~\cite{lin2017effects,lin2018population}. Here, we have provided a relationship between the dynamics of population growth rate and the distribution of single-cell growth rates. We have shown that a population starting from a single cell shows sustained oscillations in population growth due to the correlations induced by the cell size control mechanism. The timescale for the decay of these oscillations is only dependent on the single-cell growth rate distribution. As seen in \fig{decaying}, for a realistic value of single-cell growth rate variability, oscillations can be observed for as long as $40$ generations. A test tube culture of \textit{E. coli} starting from a single cell begins to saturate after about $30$ generations ($\sim 10^9$ cells/ml) making these oscillations visible at any time during the exponential growth phase\footnote{These oscillations are hidden to the optical density measurements which provide a proxy for the total mass of the population. Cell counting techniques should be used instead to detect these oscillations.}. This allows the measurement of both the steady state population growth rate $k$ and the decay rate $\lambda$. From \fig{decay_rate}, the decay rate $\lambda$ is approximately given by $\lambda \approx 29\, \sigma_\kappa^2/\bar\kappa$ which combined with \eq{k-approx} provides both the mean and the variance of single-cell growth rates, $\bar \kappa$ and $\sigma_\kappa^2$. This method for measuring the variability in single-cell growth rates is significantly easier and less biased than the direct single cell measurement. See SM~\cite{SM} for the validation of this method using the existing single-cell data from the \lq\lq mother machine" experiment from \Ref{wang2010robust}.\\

\begin{acknowledgments}
FJ gratefully acknowledges Andrea Liu, Randall Kamien, Nigel Goldenfeld, Hillel Aharoni, Jason Rocks, and Helen Ansell for their comments on the manuscript and Suckjoon Jun and Sattar Taheri-Araghi for generously sharing their data.
\end{acknowledgments}

\bibliography{ref}

\begin{thebibliography}{59}%
\makeatletter
\providecommand \@ifxundefined [1]{%
 \@ifx{#1\undefined}
}%
\providecommand \@ifnum [1]{%
 \ifnum #1\expandafter \@firstoftwo
 \else \expandafter \@secondoftwo
 \fi
}%
\providecommand \@ifx [1]{%
 \ifx #1\expandafter \@firstoftwo
 \else \expandafter \@secondoftwo
 \fi
}%
\providecommand \natexlab [1]{#1}%
\providecommand \enquote  [1]{``#1''}%
\providecommand \bibnamefont  [1]{#1}%
\providecommand \bibfnamefont [1]{#1}%
\providecommand \citenamefont [1]{#1}%
\providecommand \href@noop [0]{\@secondoftwo}%
\providecommand \href [0]{\begingroup \@sanitize@url \@href}%
\providecommand \@href[1]{\@@startlink{#1}\@@href}%
\providecommand \@@href[1]{\endgroup#1\@@endlink}%
\providecommand \@sanitize@url [0]{\catcode `\\12\catcode `\$12\catcode
  `\&12\catcode `\#12\catcode `\^12\catcode `\_12\catcode `\%12\relax}%
\providecommand \@@startlink[1]{}%
\providecommand \@@endlink[0]{}%
\providecommand \url  [0]{\begingroup\@sanitize@url \@url }%
\providecommand \@url [1]{\endgroup\@href {#1}{\urlprefix }}%
\providecommand \urlprefix  [0]{URL }%
\providecommand \Eprint [0]{\href }%
\providecommand \doibase [0]{http://dx.doi.org/}%
\providecommand \selectlanguage [0]{\@gobble}%
\providecommand \bibinfo  [0]{\@secondoftwo}%
\providecommand \bibfield  [0]{\@secondoftwo}%
\providecommand \translation [1]{[#1]}%
\providecommand \BibitemOpen [0]{}%
\providecommand \bibitemStop [0]{}%
\providecommand \bibitemNoStop [0]{.\EOS\space}%
\providecommand \EOS [0]{\spacefactor3000\relax}%
\providecommand \BibitemShut  [1]{\csname bibitem#1\endcsname}%
\let\auto@bib@innerbib\@empty
\bibitem [{\citenamefont {Jafarpour}\ \emph {et~al.}(2018)\citenamefont
  {Jafarpour}, \citenamefont {Wright}, \citenamefont {Gudjonson}, \citenamefont
  {Riebling}, \citenamefont {Dawson}, \citenamefont {Lo}, \citenamefont
  {Fiebig}, \citenamefont {Crosson}, \citenamefont {Dinner},\ and\
  \citenamefont {Iyer-Biswas}}]{jafarpour2018bridging}%
  \BibitemOpen
  \bibfield  {author} {\bibinfo {author} {\bibfnamefont {Farshid}\ \bibnamefont
  {Jafarpour}}, \bibinfo {author} {\bibfnamefont {Charles~S}\ \bibnamefont
  {Wright}}, \bibinfo {author} {\bibfnamefont {Herman}\ \bibnamefont
  {Gudjonson}}, \bibinfo {author} {\bibfnamefont {Jedidiah}\ \bibnamefont
  {Riebling}}, \bibinfo {author} {\bibfnamefont {Emma}\ \bibnamefont {Dawson}},
  \bibinfo {author} {\bibfnamefont {Klevin}\ \bibnamefont {Lo}}, \bibinfo
  {author} {\bibfnamefont {Aretha}\ \bibnamefont {Fiebig}}, \bibinfo {author}
  {\bibfnamefont {Sean}\ \bibnamefont {Crosson}}, \bibinfo {author}
  {\bibfnamefont {Aaron~R}\ \bibnamefont {Dinner}}, \ and\ \bibinfo {author}
  {\bibfnamefont {Srividya}\ \bibnamefont {Iyer-Biswas}},\ }\bibfield  {title}
  {\enquote {\bibinfo {title} {Bridging the timescales of single-cell and
  population dynamics},}\ }\href@noop {} {\bibfield  {journal} {\bibinfo
  {journal} {Physical Review X}\ }\textbf {\bibinfo {volume} {8}},\ \bibinfo
  {pages} {021007} (\bibinfo {year} {2018})}\BibitemShut {NoStop}%
\bibitem [{\citenamefont {Wang}\ \emph {et~al.}(2010)\citenamefont {Wang},
  \citenamefont {Robert}, \citenamefont {Pelletier}, \citenamefont {Dang},
  \citenamefont {Taddei}, \citenamefont {Wright},\ and\ \citenamefont
  {Jun}}]{wang2010robust}%
  \BibitemOpen
  \bibfield  {author} {\bibinfo {author} {\bibfnamefont {Ping}\ \bibnamefont
  {Wang}}, \bibinfo {author} {\bibfnamefont {Lydia}\ \bibnamefont {Robert}},
  \bibinfo {author} {\bibfnamefont {James}\ \bibnamefont {Pelletier}}, \bibinfo
  {author} {\bibfnamefont {Wei~Lien}\ \bibnamefont {Dang}}, \bibinfo {author}
  {\bibfnamefont {Francois}\ \bibnamefont {Taddei}}, \bibinfo {author}
  {\bibfnamefont {Andrew}\ \bibnamefont {Wright}}, \ and\ \bibinfo {author}
  {\bibfnamefont {Suckjoon}\ \bibnamefont {Jun}},\ }\bibfield  {title}
  {\enquote {\bibinfo {title} {Robust growth of escherichia coli},}\
  }\href@noop {} {\bibfield  {journal} {\bibinfo  {journal} {Current biology}\
  }\textbf {\bibinfo {volume} {20}},\ \bibinfo {pages} {1099--1103} (\bibinfo
  {year} {2010})}\BibitemShut {NoStop}%
\bibitem [{\citenamefont {Campos}\ \emph {et~al.}(2014)\citenamefont {Campos},
  \citenamefont {Surovtsev}, \citenamefont {Kato}, \citenamefont {Paintdakhi},
  \citenamefont {Beltran}, \citenamefont {Ebmeier},\ and\ \citenamefont
  {Jacobs-Wagner}}]{campos2014constant}%
  \BibitemOpen
  \bibfield  {author} {\bibinfo {author} {\bibfnamefont {Manuel}\ \bibnamefont
  {Campos}}, \bibinfo {author} {\bibfnamefont {Ivan~V}\ \bibnamefont
  {Surovtsev}}, \bibinfo {author} {\bibfnamefont {Setsu}\ \bibnamefont {Kato}},
  \bibinfo {author} {\bibfnamefont {Ahmad}\ \bibnamefont {Paintdakhi}},
  \bibinfo {author} {\bibfnamefont {Bruno}\ \bibnamefont {Beltran}}, \bibinfo
  {author} {\bibfnamefont {Sarah~E}\ \bibnamefont {Ebmeier}}, \ and\ \bibinfo
  {author} {\bibfnamefont {Christine}\ \bibnamefont {Jacobs-Wagner}},\
  }\bibfield  {title} {\enquote {\bibinfo {title} {A constant size extension
  drives bacterial cell size homeostasis},}\ }\href@noop {} {\bibfield
  {journal} {\bibinfo  {journal} {Cell}\ }\textbf {\bibinfo {volume} {159}},\
  \bibinfo {pages} {1433--1446} (\bibinfo {year} {2014})}\BibitemShut {NoStop}%
\bibitem [{\citenamefont {Iyer-Biswas}\ \emph
  {et~al.}(2014{\natexlab{a}})\citenamefont {Iyer-Biswas}, \citenamefont
  {Wright}, \citenamefont {Henry}, \citenamefont {Lo}, \citenamefont {Burov},
  \citenamefont {Lin}, \citenamefont {Crooks}, \citenamefont {Crosson},
  \citenamefont {Dinner},\ and\ \citenamefont {Scherer}}]{iyer2014scaling}%
  \BibitemOpen
  \bibfield  {author} {\bibinfo {author} {\bibfnamefont {Srividya}\
  \bibnamefont {Iyer-Biswas}}, \bibinfo {author} {\bibfnamefont {Charles~S}\
  \bibnamefont {Wright}}, \bibinfo {author} {\bibfnamefont {Jonathan~T}\
  \bibnamefont {Henry}}, \bibinfo {author} {\bibfnamefont {Klevin}\
  \bibnamefont {Lo}}, \bibinfo {author} {\bibfnamefont {Stanislav}\
  \bibnamefont {Burov}}, \bibinfo {author} {\bibfnamefont {Yihan}\ \bibnamefont
  {Lin}}, \bibinfo {author} {\bibfnamefont {Gavin~E}\ \bibnamefont {Crooks}},
  \bibinfo {author} {\bibfnamefont {Sean}\ \bibnamefont {Crosson}}, \bibinfo
  {author} {\bibfnamefont {Aaron~R}\ \bibnamefont {Dinner}}, \ and\ \bibinfo
  {author} {\bibfnamefont {Norbert~F}\ \bibnamefont {Scherer}},\ }\bibfield
  {title} {\enquote {\bibinfo {title} {Scaling laws governing stochastic growth
  and division of single bacterial cells},}\ }\href@noop {} {\bibfield
  {journal} {\bibinfo  {journal} {Proceedings of the National Academy of
  Sciences}\ }\textbf {\bibinfo {volume} {111}},\ \bibinfo {pages}
  {15912--15917} (\bibinfo {year} {2014}{\natexlab{a}})}\BibitemShut {NoStop}%
\bibitem [{\citenamefont {Iyer-Biswas}\ \emph
  {et~al.}(2014{\natexlab{b}})\citenamefont {Iyer-Biswas}, \citenamefont
  {Crooks}, \citenamefont {Scherer},\ and\ \citenamefont
  {Dinner}}]{iyer2014universality}%
  \BibitemOpen
  \bibfield  {author} {\bibinfo {author} {\bibfnamefont {Srividya}\
  \bibnamefont {Iyer-Biswas}}, \bibinfo {author} {\bibfnamefont {Gavin~E}\
  \bibnamefont {Crooks}}, \bibinfo {author} {\bibfnamefont {Norbert~F}\
  \bibnamefont {Scherer}}, \ and\ \bibinfo {author} {\bibfnamefont {Aaron~R}\
  \bibnamefont {Dinner}},\ }\bibfield  {title} {\enquote {\bibinfo {title}
  {Universality in stochastic exponential growth},}\ }\href@noop {} {\bibfield
  {journal} {\bibinfo  {journal} {Physical review letters}\ }\textbf {\bibinfo
  {volume} {113}},\ \bibinfo {pages} {028101} (\bibinfo {year}
  {2014}{\natexlab{b}})}\BibitemShut {NoStop}%
\bibitem [{\citenamefont {Taheri-Araghi}\ \emph {et~al.}(2015)\citenamefont
  {Taheri-Araghi}, \citenamefont {Bradde}, \citenamefont {Sauls}, \citenamefont
  {Hill}, \citenamefont {Levin}, \citenamefont {Paulsson}, \citenamefont
  {Vergassola},\ and\ \citenamefont {Jun}}]{taheri2015cell}%
  \BibitemOpen
  \bibfield  {author} {\bibinfo {author} {\bibfnamefont {Sattar}\ \bibnamefont
  {Taheri-Araghi}}, \bibinfo {author} {\bibfnamefont {Serena}\ \bibnamefont
  {Bradde}}, \bibinfo {author} {\bibfnamefont {John~T}\ \bibnamefont {Sauls}},
  \bibinfo {author} {\bibfnamefont {Norbert~S}\ \bibnamefont {Hill}}, \bibinfo
  {author} {\bibfnamefont {Petra~Anne}\ \bibnamefont {Levin}}, \bibinfo
  {author} {\bibfnamefont {Johan}\ \bibnamefont {Paulsson}}, \bibinfo {author}
  {\bibfnamefont {Massimo}\ \bibnamefont {Vergassola}}, \ and\ \bibinfo
  {author} {\bibfnamefont {Suckjoon}\ \bibnamefont {Jun}},\ }\bibfield  {title}
  {\enquote {\bibinfo {title} {Cell-size control and homeostasis in
  bacteria},}\ }\href@noop {} {\bibfield  {journal} {\bibinfo  {journal}
  {Current Biology}\ }\textbf {\bibinfo {volume} {25}},\ \bibinfo {pages}
  {385--391} (\bibinfo {year} {2015})}\BibitemShut {NoStop}%
\bibitem [{\citenamefont {Pirjol}\ \emph {et~al.}(2017)\citenamefont {Pirjol},
  \citenamefont {Jafarpour},\ and\ \citenamefont
  {Iyer-Biswas}}]{pirjol2017phenomenology}%
  \BibitemOpen
  \bibfield  {author} {\bibinfo {author} {\bibfnamefont {Dan}\ \bibnamefont
  {Pirjol}}, \bibinfo {author} {\bibfnamefont {Farshid}\ \bibnamefont
  {Jafarpour}}, \ and\ \bibinfo {author} {\bibfnamefont {Srividya}\
  \bibnamefont {Iyer-Biswas}},\ }\bibfield  {title} {\enquote {\bibinfo {title}
  {Phenomenology of stochastic exponential growth},}\ }\href@noop {} {\bibfield
   {journal} {\bibinfo  {journal} {Physical Review E}\ }\textbf {\bibinfo
  {volume} {95}},\ \bibinfo {pages} {062406} (\bibinfo {year}
  {2017})}\BibitemShut {NoStop}%
\bibitem [{\citenamefont {Di~Talia}\ \emph {et~al.}(2007)\citenamefont
  {Di~Talia}, \citenamefont {Skotheim}, \citenamefont {Bean}, \citenamefont
  {Siggia},\ and\ \citenamefont {Cross}}]{di2007effects}%
  \BibitemOpen
  \bibfield  {author} {\bibinfo {author} {\bibfnamefont {Stefano}\ \bibnamefont
  {Di~Talia}}, \bibinfo {author} {\bibfnamefont {Jan~M}\ \bibnamefont
  {Skotheim}}, \bibinfo {author} {\bibfnamefont {James~M}\ \bibnamefont
  {Bean}}, \bibinfo {author} {\bibfnamefont {Eric~D}\ \bibnamefont {Siggia}}, \
  and\ \bibinfo {author} {\bibfnamefont {Frederick~R}\ \bibnamefont {Cross}},\
  }\bibfield  {title} {\enquote {\bibinfo {title} {The effects of molecular
  noise and size control on variability in the budding yeast cell cycle},}\
  }\href@noop {} {\bibfield  {journal} {\bibinfo  {journal} {Nature}\ }\textbf
  {\bibinfo {volume} {448}},\ \bibinfo {pages} {947} (\bibinfo {year}
  {2007})}\BibitemShut {NoStop}%
\bibitem [{\citenamefont {Eun}\ \emph {et~al.}(2018)\citenamefont {Eun},
  \citenamefont {Ho}, \citenamefont {Kim}, \citenamefont {LaRussa},
  \citenamefont {Robert}, \citenamefont {Renner}, \citenamefont {Schmid},
  \citenamefont {Garner},\ and\ \citenamefont {Amir}}]{eun2018archaeal}%
  \BibitemOpen
  \bibfield  {author} {\bibinfo {author} {\bibfnamefont {Ye-Jin}\ \bibnamefont
  {Eun}}, \bibinfo {author} {\bibfnamefont {Po-Yi}\ \bibnamefont {Ho}},
  \bibinfo {author} {\bibfnamefont {Minjeong}\ \bibnamefont {Kim}}, \bibinfo
  {author} {\bibfnamefont {Salvatore}\ \bibnamefont {LaRussa}}, \bibinfo
  {author} {\bibfnamefont {Lydia}\ \bibnamefont {Robert}}, \bibinfo {author}
  {\bibfnamefont {Lars~D}\ \bibnamefont {Renner}}, \bibinfo {author}
  {\bibfnamefont {Amy}\ \bibnamefont {Schmid}}, \bibinfo {author}
  {\bibfnamefont {Ethan}\ \bibnamefont {Garner}}, \ and\ \bibinfo {author}
  {\bibfnamefont {Ariel}\ \bibnamefont {Amir}},\ }\bibfield  {title} {\enquote
  {\bibinfo {title} {Archaeal cells share common size control with bacteria
  despite noisier growth and division},}\ }\href@noop {} {\bibfield  {journal}
  {\bibinfo  {journal} {Nature microbiology}\ }\textbf {\bibinfo {volume}
  {3}},\ \bibinfo {pages} {148} (\bibinfo {year} {2018})}\BibitemShut {NoStop}%
\bibitem [{\citenamefont {Amir}(2014)}]{amir2014cell}%
  \BibitemOpen
  \bibfield  {author} {\bibinfo {author} {\bibfnamefont {Ariel}\ \bibnamefont
  {Amir}},\ }\bibfield  {title} {\enquote {\bibinfo {title} {Cell size
  regulation in bacteria},}\ }\href@noop {} {\bibfield  {journal} {\bibinfo
  {journal} {Physical Review Letters}\ }\textbf {\bibinfo {volume} {112}},\
  \bibinfo {pages} {208102} (\bibinfo {year} {2014})}\BibitemShut {NoStop}%
\bibitem [{\citenamefont {De~Martino}\ \emph {et~al.}(2016)\citenamefont
  {De~Martino}, \citenamefont {Capuani},\ and\ \citenamefont
  {De~Martino}}]{de2016growth}%
  \BibitemOpen
  \bibfield  {author} {\bibinfo {author} {\bibfnamefont {Daniele}\ \bibnamefont
  {De~Martino}}, \bibinfo {author} {\bibfnamefont {Fabrizio}\ \bibnamefont
  {Capuani}}, \ and\ \bibinfo {author} {\bibfnamefont {Andrea}\ \bibnamefont
  {De~Martino}},\ }\bibfield  {title} {\enquote {\bibinfo {title} {Growth
  against entropy in bacterial metabolism: the phenotypic trade-off behind
  empirical growth rate distributions in e. coli},}\ }\href@noop {} {\bibfield
  {journal} {\bibinfo  {journal} {Physical biology}\ }\textbf {\bibinfo
  {volume} {13}},\ \bibinfo {pages} {036005} (\bibinfo {year}
  {2016})}\BibitemShut {NoStop}%
\bibitem [{\citenamefont {Nozoe}\ \emph {et~al.}(2017)\citenamefont {Nozoe},
  \citenamefont {Kussell},\ and\ \citenamefont
  {Wakamoto}}]{nozoe2017inferring}%
  \BibitemOpen
  \bibfield  {author} {\bibinfo {author} {\bibfnamefont {Takashi}\ \bibnamefont
  {Nozoe}}, \bibinfo {author} {\bibfnamefont {Edo}\ \bibnamefont {Kussell}}, \
  and\ \bibinfo {author} {\bibfnamefont {Yuichi}\ \bibnamefont {Wakamoto}},\
  }\bibfield  {title} {\enquote {\bibinfo {title} {Inferring fitness landscapes
  and selection on phenotypic states from single-cell genealogical data},}\
  }\href@noop {} {\bibfield  {journal} {\bibinfo  {journal} {PLoS genetics}\
  }\textbf {\bibinfo {volume} {13}},\ \bibinfo {pages} {e1006653} (\bibinfo
  {year} {2017})}\BibitemShut {NoStop}%
\bibitem [{\citenamefont {Robert}\ \emph {et~al.}(2018)\citenamefont {Robert},
  \citenamefont {Ollion}, \citenamefont {Robert}, \citenamefont {Song},
  \citenamefont {Matic},\ and\ \citenamefont {Elez}}]{robert2018mutation}%
  \BibitemOpen
  \bibfield  {author} {\bibinfo {author} {\bibfnamefont {Lydia}\ \bibnamefont
  {Robert}}, \bibinfo {author} {\bibfnamefont {Jean}\ \bibnamefont {Ollion}},
  \bibinfo {author} {\bibfnamefont {Jerome}\ \bibnamefont {Robert}}, \bibinfo
  {author} {\bibfnamefont {Xiaohu}\ \bibnamefont {Song}}, \bibinfo {author}
  {\bibfnamefont {Ivan}\ \bibnamefont {Matic}}, \ and\ \bibinfo {author}
  {\bibfnamefont {Marina}\ \bibnamefont {Elez}},\ }\bibfield  {title} {\enquote
  {\bibinfo {title} {Mutation dynamics and fitness effects followed in single
  cells},}\ }\href@noop {} {\bibfield  {journal} {\bibinfo  {journal}
  {Science}\ }\textbf {\bibinfo {volume} {359}},\ \bibinfo {pages} {1283--1286}
  (\bibinfo {year} {2018})}\BibitemShut {NoStop}%
\bibitem [{\citenamefont {Lin}\ and\ \citenamefont
  {Amir}(2017)}]{lin2017effects}%
  \BibitemOpen
  \bibfield  {author} {\bibinfo {author} {\bibfnamefont {Jie}\ \bibnamefont
  {Lin}}\ and\ \bibinfo {author} {\bibfnamefont {Ariel}\ \bibnamefont {Amir}},\
  }\bibfield  {title} {\enquote {\bibinfo {title} {The effects of stochasticity
  at the single-cell level and cell size control on the population growth},}\
  }\href@noop {} {\bibfield  {journal} {\bibinfo  {journal} {Cell systems}\
  }\textbf {\bibinfo {volume} {5}},\ \bibinfo {pages} {358--367} (\bibinfo
  {year} {2017})}\BibitemShut {NoStop}%
\bibitem [{\citenamefont {Lin}\ and\ \citenamefont
  {Amir}(2018)}]{lin2018population}%
  \BibitemOpen
  \bibfield  {author} {\bibinfo {author} {\bibfnamefont {Jie}\ \bibnamefont
  {Lin}}\ and\ \bibinfo {author} {\bibfnamefont {Ariel}\ \bibnamefont {Amir}},\
  }\bibfield  {title} {\enquote {\bibinfo {title} {Population growth with
  correlated generation times at the single-cell level},}\ }\href@noop {}
  {\bibfield  {journal} {\bibinfo  {journal} {arXiv preprint arXiv:1806.02818}\
  } (\bibinfo {year} {2018})}\BibitemShut {NoStop}%
\bibitem [{\citenamefont {Ho}\ \emph {et~al.}(2018)\citenamefont {Ho},
  \citenamefont {Lin},\ and\ \citenamefont {Amir}}]{ho2018modeling}%
  \BibitemOpen
  \bibfield  {author} {\bibinfo {author} {\bibfnamefont {Po-Yi}\ \bibnamefont
  {Ho}}, \bibinfo {author} {\bibfnamefont {Jie}\ \bibnamefont {Lin}}, \ and\
  \bibinfo {author} {\bibfnamefont {Ariel}\ \bibnamefont {Amir}},\ }\bibfield
  {title} {\enquote {\bibinfo {title} {Modeling cell size regulation: From
  single-cell-level statistics to molecular mechanisms and population-level
  effects},}\ }\href@noop {} {\bibfield  {journal} {\bibinfo  {journal} {Annual
  review of biophysics}\ }\textbf {\bibinfo {volume} {47}},\ \bibinfo {pages}
  {251--271} (\bibinfo {year} {2018})}\BibitemShut {NoStop}%
\bibitem [{\citenamefont {Diekmann}\ \emph {et~al.}(1983)\citenamefont
  {Diekmann}, \citenamefont {Lauwerier}, \citenamefont {Aldenberg},\ and\
  \citenamefont {Metz}}]{diekmann1983growth}%
  \BibitemOpen
  \bibfield  {author} {\bibinfo {author} {\bibfnamefont {Odo}\ \bibnamefont
  {Diekmann}}, \bibinfo {author} {\bibfnamefont {HA}~\bibnamefont {Lauwerier}},
  \bibinfo {author} {\bibfnamefont {T}~\bibnamefont {Aldenberg}}, \ and\
  \bibinfo {author} {\bibfnamefont {JAJ}\ \bibnamefont {Metz}},\ }\bibfield
  {title} {\enquote {\bibinfo {title} {Growth, fission and the stable size
  distribution},}\ }\href@noop {} {\bibfield  {journal} {\bibinfo  {journal}
  {Journal of mathematical biology}\ }\textbf {\bibinfo {volume} {18}},\
  \bibinfo {pages} {135--148} (\bibinfo {year} {1983})}\BibitemShut {NoStop}%
\bibitem [{\citenamefont {Jun}\ and\ \citenamefont
  {Taheri-Araghi}(2015)}]{jun2015cell}%
  \BibitemOpen
  \bibfield  {author} {\bibinfo {author} {\bibfnamefont {Suckjoon}\
  \bibnamefont {Jun}}\ and\ \bibinfo {author} {\bibfnamefont {Sattar}\
  \bibnamefont {Taheri-Araghi}},\ }\bibfield  {title} {\enquote {\bibinfo
  {title} {Cell-size maintenance: universal strategy revealed},}\ }\href@noop
  {} {\bibfield  {journal} {\bibinfo  {journal} {Trends in microbiology}\
  }\textbf {\bibinfo {volume} {23}},\ \bibinfo {pages} {4--6} (\bibinfo {year}
  {2015})}\BibitemShut {NoStop}%
\bibitem [{\citenamefont {Sauls}\ \emph {et~al.}(2016)\citenamefont {Sauls},
  \citenamefont {Li},\ and\ \citenamefont {Jun}}]{sauls2016adder}%
  \BibitemOpen
  \bibfield  {author} {\bibinfo {author} {\bibfnamefont {John~T}\ \bibnamefont
  {Sauls}}, \bibinfo {author} {\bibfnamefont {Dongyang}\ \bibnamefont {Li}}, \
  and\ \bibinfo {author} {\bibfnamefont {Suckjoon}\ \bibnamefont {Jun}},\
  }\bibfield  {title} {\enquote {\bibinfo {title} {Adder and a coarse-grained
  approach to cell size homeostasis in bacteria},}\ }\href@noop {} {\bibfield
  {journal} {\bibinfo  {journal} {Current opinion in cell biology}\ }\textbf
  {\bibinfo {volume} {38}},\ \bibinfo {pages} {38--44} (\bibinfo {year}
  {2016})}\BibitemShut {NoStop}%
\bibitem [{\citenamefont {Logsdon}\ \emph {et~al.}(2017)\citenamefont
  {Logsdon}, \citenamefont {Ho}, \citenamefont {Papavinasasundaram},
  \citenamefont {Richardson}, \citenamefont {Cokol}, \citenamefont {Sassetti},
  \citenamefont {Amir},\ and\ \citenamefont {Aldridge}}]{logsdon2017parallel}%
  \BibitemOpen
  \bibfield  {author} {\bibinfo {author} {\bibfnamefont {Michelle~M}\
  \bibnamefont {Logsdon}}, \bibinfo {author} {\bibfnamefont {Po-Yi}\
  \bibnamefont {Ho}}, \bibinfo {author} {\bibfnamefont {Kadamba}\ \bibnamefont
  {Papavinasasundaram}}, \bibinfo {author} {\bibfnamefont {Kirill}\
  \bibnamefont {Richardson}}, \bibinfo {author} {\bibfnamefont {Murat}\
  \bibnamefont {Cokol}}, \bibinfo {author} {\bibfnamefont {Christopher~M}\
  \bibnamefont {Sassetti}}, \bibinfo {author} {\bibfnamefont {Ariel}\
  \bibnamefont {Amir}}, \ and\ \bibinfo {author} {\bibfnamefont {Bree~B}\
  \bibnamefont {Aldridge}},\ }\bibfield  {title} {\enquote {\bibinfo {title} {A
  parallel adder coordinates mycobacterial cell-cycle progression and cell-size
  homeostasis in the context of asymmetric growth and organization},}\
  }\href@noop {} {\bibfield  {journal} {\bibinfo  {journal} {Current Biology}\
  }\textbf {\bibinfo {volume} {27}},\ \bibinfo {pages} {3367--3374} (\bibinfo
  {year} {2017})}\BibitemShut {NoStop}%
\bibitem [{\citenamefont {Jun}\ \emph {et~al.}(2018)\citenamefont {Jun},
  \citenamefont {Si}, \citenamefont {Pugatch},\ and\ \citenamefont
  {Scott}}]{jun2018fundamental}%
  \BibitemOpen
  \bibfield  {author} {\bibinfo {author} {\bibfnamefont {Suckjoon}\
  \bibnamefont {Jun}}, \bibinfo {author} {\bibfnamefont {Fangwei}\ \bibnamefont
  {Si}}, \bibinfo {author} {\bibfnamefont {Rami}\ \bibnamefont {Pugatch}}, \
  and\ \bibinfo {author} {\bibfnamefont {Matthew}\ \bibnamefont {Scott}},\
  }\bibfield  {title} {\enquote {\bibinfo {title} {Fundamental principles in
  bacterial physiology—history, recent progress, and the future with focus on
  cell size control: a review},}\ }\href@noop {} {\bibfield  {journal}
  {\bibinfo  {journal} {Reports on Progress in Physics}\ }\textbf {\bibinfo
  {volume} {81}},\ \bibinfo {pages} {056601} (\bibinfo {year}
  {2018})}\BibitemShut {NoStop}%
\bibitem [{\citenamefont {Banerjee}\ \emph {et~al.}(2017)\citenamefont
  {Banerjee}, \citenamefont {Lo}, \citenamefont {Daddysman}, \citenamefont
  {Selewa}, \citenamefont {Kuntz}, \citenamefont {Dinner},\ and\ \citenamefont
  {Scherer}}]{banerjee2017biphasic}%
  \BibitemOpen
  \bibfield  {author} {\bibinfo {author} {\bibfnamefont {Shiladitya}\
  \bibnamefont {Banerjee}}, \bibinfo {author} {\bibfnamefont {Klevin}\
  \bibnamefont {Lo}}, \bibinfo {author} {\bibfnamefont {Matthew~K}\
  \bibnamefont {Daddysman}}, \bibinfo {author} {\bibfnamefont {Alan}\
  \bibnamefont {Selewa}}, \bibinfo {author} {\bibfnamefont {Thomas}\
  \bibnamefont {Kuntz}}, \bibinfo {author} {\bibfnamefont {Aaron~R}\
  \bibnamefont {Dinner}}, \ and\ \bibinfo {author} {\bibfnamefont {Norbert~F}\
  \bibnamefont {Scherer}},\ }\bibfield  {title} {\enquote {\bibinfo {title}
  {Biphasic growth dynamics control cell division in caulobacter crescentus},}\
  }\href@noop {} {\bibfield  {journal} {\bibinfo  {journal} {Nature
  microbiology}\ }\textbf {\bibinfo {volume} {2}},\ \bibinfo {pages} {17116}
  (\bibinfo {year} {2017})}\BibitemShut {NoStop}%
\bibitem [{\citenamefont {Si}\ \emph {et~al.}(2018)\citenamefont {Si},
  \citenamefont {Le~Treut}, \citenamefont {Sauls}, \citenamefont {Vadia},
  \citenamefont {Levin},\ and\ \citenamefont {Jun}}]{si2018mechanistic}%
  \BibitemOpen
  \bibfield  {author} {\bibinfo {author} {\bibfnamefont {Fangwei}\ \bibnamefont
  {Si}}, \bibinfo {author} {\bibfnamefont {Guillaume}\ \bibnamefont
  {Le~Treut}}, \bibinfo {author} {\bibfnamefont {John~T}\ \bibnamefont
  {Sauls}}, \bibinfo {author} {\bibfnamefont {Stephen}\ \bibnamefont {Vadia}},
  \bibinfo {author} {\bibfnamefont {Petra~Anne}\ \bibnamefont {Levin}}, \ and\
  \bibinfo {author} {\bibfnamefont {Suckjoon}\ \bibnamefont {Jun}},\ }\bibfield
   {title} {\enquote {\bibinfo {title} {Mechanistic origin of cell-size control
  and homeostasis in bacteria},}\ }\href@noop {} {\bibfield  {journal}
  {\bibinfo  {journal} {bioRxiv}\ ,\ \bibinfo {pages} {478818}} (\bibinfo
  {year} {2018})}\BibitemShut {NoStop}%
\bibitem [{SM()}]{SM}%
  \BibitemOpen
  \href@noop {} {}\bibinfo {note} {See Supplemental Material for the
  experimental validation of the model with existing data, the derivation of
  Eq.~(3), analysis of the model with stochastic division ratio, and the
  analysis of steady state population growth rate.}\BibitemShut {Stop}%
\bibitem [{\citenamefont {Guberman}\ \emph {et~al.}(2008)\citenamefont
  {Guberman}, \citenamefont {Fay}, \citenamefont {Dworkin}, \citenamefont
  {Wingreen},\ and\ \citenamefont {Gitai}}]{guberman2008psicic}%
  \BibitemOpen
  \bibfield  {author} {\bibinfo {author} {\bibfnamefont {Jonathan~M}\
  \bibnamefont {Guberman}}, \bibinfo {author} {\bibfnamefont {Allison}\
  \bibnamefont {Fay}}, \bibinfo {author} {\bibfnamefont {Jonathan}\
  \bibnamefont {Dworkin}}, \bibinfo {author} {\bibfnamefont {Ned~S}\
  \bibnamefont {Wingreen}}, \ and\ \bibinfo {author} {\bibfnamefont {Zemer}\
  \bibnamefont {Gitai}},\ }\bibfield  {title} {\enquote {\bibinfo {title}
  {Psicic: noise and asymmetry in bacterial division revealed by computational
  image analysis at sub-pixel resolution},}\ }\href@noop {} {\bibfield
  {journal} {\bibinfo  {journal} {PLoS computational biology}\ }\textbf
  {\bibinfo {volume} {4}},\ \bibinfo {pages} {e1000233} (\bibinfo {year}
  {2008})}\BibitemShut {NoStop}%
\bibitem [{\citenamefont {Cermak}\ \emph {et~al.}(2016)\citenamefont {Cermak},
  \citenamefont {Olcum}, \citenamefont {Delgado}, \citenamefont {Wasserman},
  \citenamefont {Payer}, \citenamefont {Murakami}, \citenamefont {Knudsen},
  \citenamefont {Kimmerling}, \citenamefont {Stevens}, \citenamefont {Kikuchi}
  \emph {et~al.}}]{cermak2016high}%
  \BibitemOpen
  \bibfield  {author} {\bibinfo {author} {\bibfnamefont {Nathan}\ \bibnamefont
  {Cermak}}, \bibinfo {author} {\bibfnamefont {Selim}\ \bibnamefont {Olcum}},
  \bibinfo {author} {\bibfnamefont {Francisco~Feij{\'o}}\ \bibnamefont
  {Delgado}}, \bibinfo {author} {\bibfnamefont {Steven~C}\ \bibnamefont
  {Wasserman}}, \bibinfo {author} {\bibfnamefont {Kristofor~R}\ \bibnamefont
  {Payer}}, \bibinfo {author} {\bibfnamefont {Mark~A}\ \bibnamefont
  {Murakami}}, \bibinfo {author} {\bibfnamefont {Scott~M}\ \bibnamefont
  {Knudsen}}, \bibinfo {author} {\bibfnamefont {Robert~J}\ \bibnamefont
  {Kimmerling}}, \bibinfo {author} {\bibfnamefont {Mark~M}\ \bibnamefont
  {Stevens}}, \bibinfo {author} {\bibfnamefont {Yuki}\ \bibnamefont {Kikuchi}},
   \emph {et~al.},\ }\bibfield  {title} {\enquote {\bibinfo {title}
  {High-throughput measurement of single-cell growth rates using serial
  microfluidic mass sensor arrays},}\ }\href@noop {} {\bibfield  {journal}
  {\bibinfo  {journal} {Nature biotechnology}\ }\textbf {\bibinfo {volume}
  {34}},\ \bibinfo {pages} {1052} (\bibinfo {year} {2016})}\BibitemShut
  {NoStop}%
\bibitem [{\citenamefont {Wallden}\ \emph {et~al.}(2016)\citenamefont
  {Wallden}, \citenamefont {Fange}, \citenamefont {Lundius}, \citenamefont
  {Baltekin},\ and\ \citenamefont {Elf}}]{wallden2016synchronization}%
  \BibitemOpen
  \bibfield  {author} {\bibinfo {author} {\bibfnamefont {Mats}\ \bibnamefont
  {Wallden}}, \bibinfo {author} {\bibfnamefont {David}\ \bibnamefont {Fange}},
  \bibinfo {author} {\bibfnamefont {Ebba~Gregorsson}\ \bibnamefont {Lundius}},
  \bibinfo {author} {\bibfnamefont {{\"O}zden}\ \bibnamefont {Baltekin}}, \
  and\ \bibinfo {author} {\bibfnamefont {Johan}\ \bibnamefont {Elf}},\
  }\bibfield  {title} {\enquote {\bibinfo {title} {The synchronization of
  replication and division cycles in individual e. coli cells},}\ }\href@noop
  {} {\bibfield  {journal} {\bibinfo  {journal} {Cell}\ }\textbf {\bibinfo
  {volume} {166}},\ \bibinfo {pages} {729--739} (\bibinfo {year}
  {2016})}\BibitemShut {NoStop}%
\bibitem [{\citenamefont {Kennard}\ \emph {et~al.}(2016)\citenamefont
  {Kennard}, \citenamefont {Osella}, \citenamefont {Javer}, \citenamefont
  {Grilli}, \citenamefont {Nghe}, \citenamefont {Tans}, \citenamefont
  {Cicuta},\ and\ \citenamefont {Lagomarsino}}]{kennard2016individuality}%
  \BibitemOpen
  \bibfield  {author} {\bibinfo {author} {\bibfnamefont {Andrew~S}\
  \bibnamefont {Kennard}}, \bibinfo {author} {\bibfnamefont {Matteo}\
  \bibnamefont {Osella}}, \bibinfo {author} {\bibfnamefont {Avelino}\
  \bibnamefont {Javer}}, \bibinfo {author} {\bibfnamefont {Jacopo}\
  \bibnamefont {Grilli}}, \bibinfo {author} {\bibfnamefont {Philippe}\
  \bibnamefont {Nghe}}, \bibinfo {author} {\bibfnamefont {Sander~J}\
  \bibnamefont {Tans}}, \bibinfo {author} {\bibfnamefont {Pietro}\ \bibnamefont
  {Cicuta}}, \ and\ \bibinfo {author} {\bibfnamefont {Marco~Cosentino}\
  \bibnamefont {Lagomarsino}},\ }\bibfield  {title} {\enquote {\bibinfo {title}
  {Individuality and universality in the growth-division laws of single e. coli
  cells},}\ }\href@noop {} {\bibfield  {journal} {\bibinfo  {journal} {Physical
  Review E}\ }\textbf {\bibinfo {volume} {93}},\ \bibinfo {pages} {012408}
  (\bibinfo {year} {2016})}\BibitemShut {NoStop}%
\bibitem [{\citenamefont {Grilli}\ \emph {et~al.}(2018)\citenamefont {Grilli},
  \citenamefont {Cadart}, \citenamefont {Micali}, \citenamefont {Osella},\ and\
  \citenamefont {Lagomarsino}}]{grilli2018empirical}%
  \BibitemOpen
  \bibfield  {author} {\bibinfo {author} {\bibfnamefont {Jacopo}\ \bibnamefont
  {Grilli}}, \bibinfo {author} {\bibfnamefont {Clotilde}\ \bibnamefont
  {Cadart}}, \bibinfo {author} {\bibfnamefont {Gabriele}\ \bibnamefont
  {Micali}}, \bibinfo {author} {\bibfnamefont {Matteo}\ \bibnamefont {Osella}},
  \ and\ \bibinfo {author} {\bibfnamefont {Marco~Cosentino}\ \bibnamefont
  {Lagomarsino}},\ }\bibfield  {title} {\enquote {\bibinfo {title} {The
  empirical fluctuation pattern of e. coli division control},}\ }\href@noop {}
  {\bibfield  {journal} {\bibinfo  {journal} {Frontiers in microbiology}\
  }\textbf {\bibinfo {volume} {9}} (\bibinfo {year} {2018})}\BibitemShut
  {NoStop}%
\bibitem [{\citenamefont {Powell}(1956)}]{powell1956growth}%
  \BibitemOpen
  \bibfield  {author} {\bibinfo {author} {\bibfnamefont {EO}~\bibnamefont
  {Powell}},\ }\bibfield  {title} {\enquote {\bibinfo {title} {Growth rate and
  generation time of bacteria, with special reference to continuous culture},}\
  }\href@noop {} {\bibfield  {journal} {\bibinfo  {journal} {Microbiology}\
  }\textbf {\bibinfo {volume} {15}},\ \bibinfo {pages} {492--511} (\bibinfo
  {year} {1956})}\BibitemShut {NoStop}%
\bibitem [{\citenamefont {Rochman}\ \emph {et~al.}(2018)\citenamefont
  {Rochman}, \citenamefont {Popescu},\ and\ \citenamefont
  {Sun}}]{rochman2018ergodicity}%
  \BibitemOpen
  \bibfield  {author} {\bibinfo {author} {\bibfnamefont {Nash~D}\ \bibnamefont
  {Rochman}}, \bibinfo {author} {\bibfnamefont {Dan~M}\ \bibnamefont
  {Popescu}}, \ and\ \bibinfo {author} {\bibfnamefont {Sean~X}\ \bibnamefont
  {Sun}},\ }\bibfield  {title} {\enquote {\bibinfo {title} {Ergodicity, hidden
  bias and the growth rate gain},}\ }\href@noop {} {\bibfield  {journal}
  {\bibinfo  {journal} {Physical biology}\ }\textbf {\bibinfo {volume} {15}},\
  \bibinfo {pages} {036006} (\bibinfo {year} {2018})}\BibitemShut {NoStop}%
\bibitem [{\citenamefont {Susman}\ \emph {et~al.}(2018)\citenamefont {Susman},
  \citenamefont {Kohram}, \citenamefont {Vashistha}, \citenamefont {Nechleba},
  \citenamefont {Salman},\ and\ \citenamefont
  {Brenner}}]{susman2018individuality}%
  \BibitemOpen
  \bibfield  {author} {\bibinfo {author} {\bibfnamefont {Lee}\ \bibnamefont
  {Susman}}, \bibinfo {author} {\bibfnamefont {Maryam}\ \bibnamefont {Kohram}},
  \bibinfo {author} {\bibfnamefont {Harsh}\ \bibnamefont {Vashistha}}, \bibinfo
  {author} {\bibfnamefont {Jeffrey~T}\ \bibnamefont {Nechleba}}, \bibinfo
  {author} {\bibfnamefont {Hanna}\ \bibnamefont {Salman}}, \ and\ \bibinfo
  {author} {\bibfnamefont {Naama}\ \bibnamefont {Brenner}},\ }\bibfield
  {title} {\enquote {\bibinfo {title} {Individuality and slow dynamics in
  bacterial growth homeostasis},}\ }\href@noop {} {\bibfield  {journal}
  {\bibinfo  {journal} {Proceedings of the National Academy of Sciences}\ ,\
  \bibinfo {pages} {201615526}} (\bibinfo {year} {2018})}\BibitemShut {NoStop}%
\bibitem [{\citenamefont {M'Kendrick}(1925)}]{m1925applications}%
  \BibitemOpen
  \bibfield  {author} {\bibinfo {author} {\bibfnamefont {AG}~\bibnamefont
  {M'Kendrick}},\ }\bibfield  {title} {\enquote {\bibinfo {title} {Applications
  of mathematics to medical problems},}\ }\href@noop {} {\bibfield  {journal}
  {\bibinfo  {journal} {Proceedings of the Edinburgh Mathematical Society}\
  }\textbf {\bibinfo {volume} {44}},\ \bibinfo {pages} {98--130} (\bibinfo
  {year} {1925})}\BibitemShut {NoStop}%
\bibitem [{\citenamefont {Von~Foerster}\ and\ \citenamefont
  {Stohlman}(1959)}]{von1959kinetics}%
  \BibitemOpen
  \bibfield  {author} {\bibinfo {author} {\bibfnamefont {H}~\bibnamefont
  {Von~Foerster}}\ and\ \bibinfo {author} {\bibfnamefont {F}~\bibnamefont
  {Stohlman}},\ }\bibfield  {title} {\enquote {\bibinfo {title} {The kinetics
  of cellular proliferation},}\ }\href@noop {} {\bibfield  {journal} {\bibinfo
  {journal} {Grune \& Stratton}\ }\textbf {\bibinfo {volume} {3}} (\bibinfo
  {year} {1959})}\BibitemShut {NoStop}%
\bibitem [{\citenamefont {Painter}\ and\ \citenamefont
  {Marr}(1968)}]{painter1968mathematics}%
  \BibitemOpen
  \bibfield  {author} {\bibinfo {author} {\bibfnamefont {PR}~\bibnamefont
  {Painter}}\ and\ \bibinfo {author} {\bibfnamefont {AG}~\bibnamefont {Marr}},\
  }\bibfield  {title} {\enquote {\bibinfo {title} {Mathematics of microbial
  populations},}\ }\href@noop {} {\bibfield  {journal} {\bibinfo  {journal}
  {Annual Reviews in Microbiology}\ }\textbf {\bibinfo {volume} {22}},\
  \bibinfo {pages} {519--548} (\bibinfo {year} {1968})}\BibitemShut {NoStop}%
\bibitem [{\citenamefont {Perthame}(2006)}]{perthame2006transport}%
  \BibitemOpen
  \bibfield  {author} {\bibinfo {author} {\bibfnamefont {Beno{\^\i}t}\
  \bibnamefont {Perthame}},\ }\href@noop {} {\emph {\bibinfo {title} {Transport
  equations in biology}}}\ (\bibinfo  {publisher} {Springer Science \& Business
  Media},\ \bibinfo {year} {2006})\BibitemShut {NoStop}%
\bibitem [{\citenamefont {Rubin}\ and\ \citenamefont
  {Riznichenko}(2016)}]{rubin2016mathematical}%
  \BibitemOpen
  \bibfield  {author} {\bibinfo {author} {\bibfnamefont {Andrew}\ \bibnamefont
  {Rubin}}\ and\ \bibinfo {author} {\bibfnamefont {Galina}\ \bibnamefont
  {Riznichenko}},\ }\href@noop {} {\emph {\bibinfo {title} {Mathematical
  biophysics}}}\ (\bibinfo  {publisher} {Springer},\ \bibinfo {year}
  {2016})\BibitemShut {NoStop}%
\bibitem [{\citenamefont {Wakamoto}\ \emph {et~al.}(2005)\citenamefont
  {Wakamoto}, \citenamefont {Ramsden},\ and\ \citenamefont
  {Yasuda}}]{wakamoto2005single}%
  \BibitemOpen
  \bibfield  {author} {\bibinfo {author} {\bibfnamefont {Yuichi}\ \bibnamefont
  {Wakamoto}}, \bibinfo {author} {\bibfnamefont {Jeremy}\ \bibnamefont
  {Ramsden}}, \ and\ \bibinfo {author} {\bibfnamefont {Kenji}\ \bibnamefont
  {Yasuda}},\ }\bibfield  {title} {\enquote {\bibinfo {title} {Single-cell
  growth and division dynamics showing epigenetic correlations},}\ }\href@noop
  {} {\bibfield  {journal} {\bibinfo  {journal} {Analyst}\ }\textbf {\bibinfo
  {volume} {130}},\ \bibinfo {pages} {311--317} (\bibinfo {year}
  {2005})}\BibitemShut {NoStop}%
\bibitem [{\citenamefont {Siegal-Gaskins}\ and\ \citenamefont
  {Crosson}(2008)}]{siegal2008tightly}%
  \BibitemOpen
  \bibfield  {author} {\bibinfo {author} {\bibfnamefont {Dan}\ \bibnamefont
  {Siegal-Gaskins}}\ and\ \bibinfo {author} {\bibfnamefont {Sean}\ \bibnamefont
  {Crosson}},\ }\bibfield  {title} {\enquote {\bibinfo {title} {Tightly
  regulated and heritable division control in single bacterial cells},}\
  }\href@noop {} {\bibfield  {journal} {\bibinfo  {journal} {Biophysical
  Journal}\ }\textbf {\bibinfo {volume} {95}},\ \bibinfo {pages} {2063--2072}
  (\bibinfo {year} {2008})}\BibitemShut {NoStop}%
\bibitem [{\citenamefont {Sliusarenko}\ \emph {et~al.}(2011)\citenamefont
  {Sliusarenko}, \citenamefont {Heinritz}, \citenamefont {Emonet},\ and\
  \citenamefont {Jacobs-Wagner}}]{sliusarenko2011high}%
  \BibitemOpen
  \bibfield  {author} {\bibinfo {author} {\bibfnamefont {Oleksii}\ \bibnamefont
  {Sliusarenko}}, \bibinfo {author} {\bibfnamefont {Jennifer}\ \bibnamefont
  {Heinritz}}, \bibinfo {author} {\bibfnamefont {Thierry}\ \bibnamefont
  {Emonet}}, \ and\ \bibinfo {author} {\bibfnamefont {Christine}\ \bibnamefont
  {Jacobs-Wagner}},\ }\bibfield  {title} {\enquote {\bibinfo {title}
  {High-throughput, subpixel precision analysis of bacterial morphogenesis and
  intracellular spatio-temporal dynamics},}\ }\href@noop {} {\bibfield
  {journal} {\bibinfo  {journal} {Molecular microbiology}\ }\textbf {\bibinfo
  {volume} {80}},\ \bibinfo {pages} {612--627} (\bibinfo {year}
  {2011})}\BibitemShut {NoStop}%
\bibitem [{\citenamefont {Young}\ \emph {et~al.}(2012)\citenamefont {Young},
  \citenamefont {Locke}, \citenamefont {Altinok}, \citenamefont {Rosenfeld},
  \citenamefont {Bacarian}, \citenamefont {Swain}, \citenamefont {Mjolsness},\
  and\ \citenamefont {Elowitz}}]{young2012measuring}%
  \BibitemOpen
  \bibfield  {author} {\bibinfo {author} {\bibfnamefont {Jonathan~W}\
  \bibnamefont {Young}}, \bibinfo {author} {\bibfnamefont {James~CW}\
  \bibnamefont {Locke}}, \bibinfo {author} {\bibfnamefont {Alphan}\
  \bibnamefont {Altinok}}, \bibinfo {author} {\bibfnamefont {Nitzan}\
  \bibnamefont {Rosenfeld}}, \bibinfo {author} {\bibfnamefont {Tigran}\
  \bibnamefont {Bacarian}}, \bibinfo {author} {\bibfnamefont {Peter~S}\
  \bibnamefont {Swain}}, \bibinfo {author} {\bibfnamefont {Eric}\ \bibnamefont
  {Mjolsness}}, \ and\ \bibinfo {author} {\bibfnamefont {Michael~B}\
  \bibnamefont {Elowitz}},\ }\bibfield  {title} {\enquote {\bibinfo {title}
  {Measuring single-cell gene expression dynamics in bacteria using
  fluorescence time-lapse microscopy},}\ }\href@noop {} {\bibfield  {journal}
  {\bibinfo  {journal} {Nature protocols}\ }\textbf {\bibinfo {volume} {7}},\
  \bibinfo {pages} {80} (\bibinfo {year} {2012})}\BibitemShut {NoStop}%
\bibitem [{\citenamefont {Salman}\ \emph {et~al.}(2012)\citenamefont {Salman},
  \citenamefont {Brenner}, \citenamefont {Tung}, \citenamefont {Elyahu},
  \citenamefont {Stolovicki}, \citenamefont {Moore}, \citenamefont
  {Libchaber},\ and\ \citenamefont {Braun}}]{salman2012universal}%
  \BibitemOpen
  \bibfield  {author} {\bibinfo {author} {\bibfnamefont {Hanna}\ \bibnamefont
  {Salman}}, \bibinfo {author} {\bibfnamefont {Naama}\ \bibnamefont {Brenner}},
  \bibinfo {author} {\bibfnamefont {Chih-kuan}\ \bibnamefont {Tung}}, \bibinfo
  {author} {\bibfnamefont {Noa}\ \bibnamefont {Elyahu}}, \bibinfo {author}
  {\bibfnamefont {Elad}\ \bibnamefont {Stolovicki}}, \bibinfo {author}
  {\bibfnamefont {Lindsay}\ \bibnamefont {Moore}}, \bibinfo {author}
  {\bibfnamefont {Albert}\ \bibnamefont {Libchaber}}, \ and\ \bibinfo {author}
  {\bibfnamefont {Erez}\ \bibnamefont {Braun}},\ }\bibfield  {title} {\enquote
  {\bibinfo {title} {Universal protein fluctuations in populations of
  microorganisms},}\ }\href@noop {} {\bibfield  {journal} {\bibinfo  {journal}
  {Physical review letters}\ }\textbf {\bibinfo {volume} {108}},\ \bibinfo
  {pages} {238105} (\bibinfo {year} {2012})}\BibitemShut {NoStop}%
\bibitem [{\citenamefont {Lambert}\ and\ \citenamefont
  {Kussell}(2015)}]{lambert2015quantifying}%
  \BibitemOpen
  \bibfield  {author} {\bibinfo {author} {\bibfnamefont {Guillaume}\
  \bibnamefont {Lambert}}\ and\ \bibinfo {author} {\bibfnamefont {Edo}\
  \bibnamefont {Kussell}},\ }\bibfield  {title} {\enquote {\bibinfo {title}
  {Quantifying selective pressures driving bacterial evolution using lineage
  analysis},}\ }\href@noop {} {\bibfield  {journal} {\bibinfo  {journal}
  {Physical review X}\ }\textbf {\bibinfo {volume} {5}},\ \bibinfo {pages}
  {011016} (\bibinfo {year} {2015})}\BibitemShut {NoStop}%
\bibitem [{\citenamefont {Schaechter}\ \emph {et~al.}(1962)\citenamefont
  {Schaechter}, \citenamefont {Williamson}, \citenamefont {Jun},\ and\
  \citenamefont {Koch}}]{schaechter1962growth}%
  \BibitemOpen
  \bibfield  {author} {\bibinfo {author} {\bibfnamefont {M}~\bibnamefont
  {Schaechter}}, \bibinfo {author} {\bibfnamefont {Joan~P}\ \bibnamefont
  {Williamson}}, \bibinfo {author} {\bibfnamefont {JR~Hood}\ \bibnamefont
  {Jun}}, \ and\ \bibinfo {author} {\bibfnamefont {Arthur~L}\ \bibnamefont
  {Koch}},\ }\bibfield  {title} {\enquote {\bibinfo {title} {Growth, cell and
  nuclear divisions in some bacteria},}\ }\href@noop {} {\bibfield  {journal}
  {\bibinfo  {journal} {Microbiology}\ }\textbf {\bibinfo {volume} {29}},\
  \bibinfo {pages} {421--434} (\bibinfo {year} {1962})}\BibitemShut {NoStop}%
\bibitem [{\citenamefont {Koch}\ and\ \citenamefont
  {Schaechter}(1962)}]{koch1962model}%
  \BibitemOpen
  \bibfield  {author} {\bibinfo {author} {\bibfnamefont {AL}~\bibnamefont
  {Koch}}\ and\ \bibinfo {author} {\bibfnamefont {M}~\bibnamefont
  {Schaechter}},\ }\bibfield  {title} {\enquote {\bibinfo {title} {A model for
  statistics of the cell division process},}\ }\href@noop {} {\bibfield
  {journal} {\bibinfo  {journal} {Microbiology}\ }\textbf {\bibinfo {volume}
  {29}},\ \bibinfo {pages} {435--454} (\bibinfo {year} {1962})}\BibitemShut
  {NoStop}%
\bibitem [{\citenamefont {Anderson}\ \emph {et~al.}(1969)\citenamefont
  {Anderson}, \citenamefont {Bell}, \citenamefont {Petersen},\ and\
  \citenamefont {Tobey}}]{anderson1969cell}%
  \BibitemOpen
  \bibfield  {author} {\bibinfo {author} {\bibfnamefont {EC}~\bibnamefont
  {Anderson}}, \bibinfo {author} {\bibfnamefont {GI}~\bibnamefont {Bell}},
  \bibinfo {author} {\bibfnamefont {DF}~\bibnamefont {Petersen}}, \ and\
  \bibinfo {author} {\bibfnamefont {RA}~\bibnamefont {Tobey}},\ }\bibfield
  {title} {\enquote {\bibinfo {title} {Cell growth and division: Iv.
  determination of volume growth rate and division probability},}\ }\href@noop
  {} {\bibfield  {journal} {\bibinfo  {journal} {Biophysical journal}\ }\textbf
  {\bibinfo {volume} {9}},\ \bibinfo {pages} {246} (\bibinfo {year}
  {1969})}\BibitemShut {NoStop}%
\bibitem [{\citenamefont {Fantes}\ \emph {et~al.}(1975)\citenamefont {Fantes},
  \citenamefont {Grant}, \citenamefont {Pritchard}, \citenamefont {Sudbery},\
  and\ \citenamefont {Wheals}}]{fantes1975regulation}%
  \BibitemOpen
  \bibfield  {author} {\bibinfo {author} {\bibfnamefont {Peter~A}\ \bibnamefont
  {Fantes}}, \bibinfo {author} {\bibfnamefont {WD}~\bibnamefont {Grant}},
  \bibinfo {author} {\bibfnamefont {RH}~\bibnamefont {Pritchard}}, \bibinfo
  {author} {\bibfnamefont {PE}~\bibnamefont {Sudbery}}, \ and\ \bibinfo
  {author} {\bibfnamefont {AE}~\bibnamefont {Wheals}},\ }\bibfield  {title}
  {\enquote {\bibinfo {title} {The regulation of cell size and the control of
  mitosis},}\ }\href@noop {} {\bibfield  {journal} {\bibinfo  {journal}
  {Journal of theoretical biology}\ }\textbf {\bibinfo {volume} {50}},\
  \bibinfo {pages} {213--244} (\bibinfo {year} {1975})}\BibitemShut {NoStop}%
\bibitem [{\citenamefont {Jorgensen}\ and\ \citenamefont
  {Tyers}(2004)}]{jorgensen2004cells}%
  \BibitemOpen
  \bibfield  {author} {\bibinfo {author} {\bibfnamefont {Paul}\ \bibnamefont
  {Jorgensen}}\ and\ \bibinfo {author} {\bibfnamefont {Mike}\ \bibnamefont
  {Tyers}},\ }\bibfield  {title} {\enquote {\bibinfo {title} {How cells
  coordinate growth and division},}\ }\href@noop {} {\bibfield  {journal}
  {\bibinfo  {journal} {Current Biology}\ }\textbf {\bibinfo {volume} {14}},\
  \bibinfo {pages} {R1014--R1027} (\bibinfo {year} {2004})}\BibitemShut
  {NoStop}%
\bibitem [{\citenamefont {Chien}\ \emph {et~al.}(2012)\citenamefont {Chien},
  \citenamefont {Hill},\ and\ \citenamefont {Levin}}]{chien2012cell}%
  \BibitemOpen
  \bibfield  {author} {\bibinfo {author} {\bibfnamefont {An-Chun}\ \bibnamefont
  {Chien}}, \bibinfo {author} {\bibfnamefont {Norbert~S}\ \bibnamefont {Hill}},
  \ and\ \bibinfo {author} {\bibfnamefont {Petra~Anne}\ \bibnamefont {Levin}},\
  }\bibfield  {title} {\enquote {\bibinfo {title} {Cell size control in
  bacteria},}\ }\href@noop {} {\bibfield  {journal} {\bibinfo  {journal}
  {Current biology}\ }\textbf {\bibinfo {volume} {22}},\ \bibinfo {pages}
  {R340--R349} (\bibinfo {year} {2012})}\BibitemShut {NoStop}%
\bibitem [{\citenamefont {Turner}\ \emph {et~al.}(2012)\citenamefont {Turner},
  \citenamefont {Ewald},\ and\ \citenamefont {Skotheim}}]{turner2012cell}%
  \BibitemOpen
  \bibfield  {author} {\bibinfo {author} {\bibfnamefont {Jonathan~J}\
  \bibnamefont {Turner}}, \bibinfo {author} {\bibfnamefont {Jennifer~C}\
  \bibnamefont {Ewald}}, \ and\ \bibinfo {author} {\bibfnamefont {Jan~M}\
  \bibnamefont {Skotheim}},\ }\bibfield  {title} {\enquote {\bibinfo {title}
  {Cell size control in yeast},}\ }\href@noop {} {\bibfield  {journal}
  {\bibinfo  {journal} {Current biology}\ }\textbf {\bibinfo {volume} {22}},\
  \bibinfo {pages} {R350--R359} (\bibinfo {year} {2012})}\BibitemShut {NoStop}%
\bibitem [{\citenamefont {Lloyd}(2013)}]{lloyd2013regulation}%
  \BibitemOpen
  \bibfield  {author} {\bibinfo {author} {\bibfnamefont {Alison~C}\
  \bibnamefont {Lloyd}},\ }\bibfield  {title} {\enquote {\bibinfo {title} {The
  regulation of cell size},}\ }\href@noop {} {\bibfield  {journal} {\bibinfo
  {journal} {Cell}\ }\textbf {\bibinfo {volume} {154}},\ \bibinfo {pages}
  {1194--1205} (\bibinfo {year} {2013})}\BibitemShut {NoStop}%
\bibitem [{\citenamefont {Osella}\ \emph {et~al.}(2014)\citenamefont {Osella},
  \citenamefont {Nugent},\ and\ \citenamefont
  {Lagomarsino}}]{osella2014concerted}%
  \BibitemOpen
  \bibfield  {author} {\bibinfo {author} {\bibfnamefont {Matteo}\ \bibnamefont
  {Osella}}, \bibinfo {author} {\bibfnamefont {Eileen}\ \bibnamefont {Nugent}},
  \ and\ \bibinfo {author} {\bibfnamefont {Marco~Cosentino}\ \bibnamefont
  {Lagomarsino}},\ }\bibfield  {title} {\enquote {\bibinfo {title} {Concerted
  control of escherichia coli cell division},}\ }\href@noop {} {\bibfield
  {journal} {\bibinfo  {journal} {Proceedings of the National Academy of
  Sciences}\ ,\ \bibinfo {pages} {201313715}} (\bibinfo {year}
  {2014})}\BibitemShut {NoStop}%
\bibitem [{\citenamefont {Robert}\ \emph {et~al.}(2014)\citenamefont {Robert},
  \citenamefont {Hoffmann}, \citenamefont {Krell}, \citenamefont {Aymerich},
  \citenamefont {Robert},\ and\ \citenamefont {Doumic}}]{robert2014division}%
  \BibitemOpen
  \bibfield  {author} {\bibinfo {author} {\bibfnamefont {Lydia}\ \bibnamefont
  {Robert}}, \bibinfo {author} {\bibfnamefont {Marc}\ \bibnamefont {Hoffmann}},
  \bibinfo {author} {\bibfnamefont {Nathalie}\ \bibnamefont {Krell}}, \bibinfo
  {author} {\bibfnamefont {St{\'e}phane}\ \bibnamefont {Aymerich}}, \bibinfo
  {author} {\bibfnamefont {J{\'e}r{\^o}me}\ \bibnamefont {Robert}}, \ and\
  \bibinfo {author} {\bibfnamefont {Marie}\ \bibnamefont {Doumic}},\ }\bibfield
   {title} {\enquote {\bibinfo {title} {Division in escherichia coli is
  triggered by a size-sensing rather than a timing mechanism},}\ }\href@noop {}
  {\bibfield  {journal} {\bibinfo  {journal} {BMC biology}\ }\textbf {\bibinfo
  {volume} {12}},\ \bibinfo {pages} {17} (\bibinfo {year} {2014})}\BibitemShut
  {NoStop}%
\bibitem [{\citenamefont {Marantan}\ and\ \citenamefont
  {Amir}(2016)}]{marantan2016stochastic}%
  \BibitemOpen
  \bibfield  {author} {\bibinfo {author} {\bibfnamefont {Andrew}\ \bibnamefont
  {Marantan}}\ and\ \bibinfo {author} {\bibfnamefont {Ariel}\ \bibnamefont
  {Amir}},\ }\bibfield  {title} {\enquote {\bibinfo {title} {Stochastic
  modeling of cell growth with symmetric or asymmetric division},}\ }\href@noop
  {} {\bibfield  {journal} {\bibinfo  {journal} {Physical Review E}\ }\textbf
  {\bibinfo {volume} {94}},\ \bibinfo {pages} {012405} (\bibinfo {year}
  {2016})}\BibitemShut {NoStop}%
\bibitem [{\citenamefont {Grilli}\ \emph {et~al.}(2017)\citenamefont {Grilli},
  \citenamefont {Osella}, \citenamefont {Kennard},\ and\ \citenamefont
  {Lagomarsino}}]{grilli2017relevant}%
  \BibitemOpen
  \bibfield  {author} {\bibinfo {author} {\bibfnamefont {Jacopo}\ \bibnamefont
  {Grilli}}, \bibinfo {author} {\bibfnamefont {Matteo}\ \bibnamefont {Osella}},
  \bibinfo {author} {\bibfnamefont {Andrew~S}\ \bibnamefont {Kennard}}, \ and\
  \bibinfo {author} {\bibfnamefont {Marco~Cosentino}\ \bibnamefont
  {Lagomarsino}},\ }\bibfield  {title} {\enquote {\bibinfo {title} {Relevant
  parameters in models of cell division control},}\ }\href@noop {} {\bibfield
  {journal} {\bibinfo  {journal} {Physical Review E}\ }\textbf {\bibinfo
  {volume} {95}},\ \bibinfo {pages} {032411} (\bibinfo {year}
  {2017})}\BibitemShut {NoStop}%
\bibitem [{\citenamefont {Cadart}\ \emph {et~al.}(2018)\citenamefont {Cadart},
  \citenamefont {Monnier}, \citenamefont {Grilli}, \citenamefont {S{\'a}ez},
  \citenamefont {Srivastava}, \citenamefont {Attia}, \citenamefont {Terriac},
  \citenamefont {Baum}, \citenamefont {Cosentino-Lagomarsino},\ and\
  \citenamefont {Piel}}]{cadart2018size}%
  \BibitemOpen
  \bibfield  {author} {\bibinfo {author} {\bibfnamefont {Clotilde}\
  \bibnamefont {Cadart}}, \bibinfo {author} {\bibfnamefont {Sylvain}\
  \bibnamefont {Monnier}}, \bibinfo {author} {\bibfnamefont {Jacopo}\
  \bibnamefont {Grilli}}, \bibinfo {author} {\bibfnamefont {Pablo~J}\
  \bibnamefont {S{\'a}ez}}, \bibinfo {author} {\bibfnamefont {Nishit}\
  \bibnamefont {Srivastava}}, \bibinfo {author} {\bibfnamefont {Rafaele}\
  \bibnamefont {Attia}}, \bibinfo {author} {\bibfnamefont {Emmanuel}\
  \bibnamefont {Terriac}}, \bibinfo {author} {\bibfnamefont {Buzz}\
  \bibnamefont {Baum}}, \bibinfo {author} {\bibfnamefont {Marco}\ \bibnamefont
  {Cosentino-Lagomarsino}}, \ and\ \bibinfo {author} {\bibfnamefont {Matthieu}\
  \bibnamefont {Piel}},\ }\bibfield  {title} {\enquote {\bibinfo {title} {Size
  control in mammalian cells involves modulation of both growth rate and cell
  cycle duration},}\ }\href@noop {} {\bibfield  {journal} {\bibinfo  {journal}
  {Nature communications}\ }\textbf {\bibinfo {volume} {9}},\ \bibinfo {pages}
  {3275} (\bibinfo {year} {2018})}\BibitemShut {NoStop}%
\bibitem [{\citenamefont {Hashimoto}\ \emph {et~al.}(2016)\citenamefont
  {Hashimoto}, \citenamefont {Nozoe}, \citenamefont {Nakaoka}, \citenamefont
  {Okura}, \citenamefont {Akiyoshi}, \citenamefont {Kaneko}, \citenamefont
  {Kussell},\ and\ \citenamefont {Wakamoto}}]{hashimoto2016noise}%
  \BibitemOpen
  \bibfield  {author} {\bibinfo {author} {\bibfnamefont {Mikihiro}\
  \bibnamefont {Hashimoto}}, \bibinfo {author} {\bibfnamefont {Takashi}\
  \bibnamefont {Nozoe}}, \bibinfo {author} {\bibfnamefont {Hidenori}\
  \bibnamefont {Nakaoka}}, \bibinfo {author} {\bibfnamefont {Reiko}\
  \bibnamefont {Okura}}, \bibinfo {author} {\bibfnamefont {Sayo}\ \bibnamefont
  {Akiyoshi}}, \bibinfo {author} {\bibfnamefont {Kunihiko}\ \bibnamefont
  {Kaneko}}, \bibinfo {author} {\bibfnamefont {Edo}\ \bibnamefont {Kussell}}, \
  and\ \bibinfo {author} {\bibfnamefont {Yuichi}\ \bibnamefont {Wakamoto}},\
  }\bibfield  {title} {\enquote {\bibinfo {title} {Noise-driven growth rate
  gain in clonal cellular populations},}\ }\href@noop {} {\bibfield  {journal}
  {\bibinfo  {journal} {Proceedings of the National Academy of Sciences}\
  }\textbf {\bibinfo {volume} {113}},\ \bibinfo {pages} {3251--3256} (\bibinfo
  {year} {2016})}\BibitemShut {NoStop}%
\bibitem [{\citenamefont {Cerulus}\ \emph {et~al.}(2016)\citenamefont
  {Cerulus}, \citenamefont {New}, \citenamefont {Pougach},\ and\ \citenamefont
  {Verstrepen}}]{cerulus2016noise}%
  \BibitemOpen
  \bibfield  {author} {\bibinfo {author} {\bibfnamefont {Bram}\ \bibnamefont
  {Cerulus}}, \bibinfo {author} {\bibfnamefont {Aaron~M}\ \bibnamefont {New}},
  \bibinfo {author} {\bibfnamefont {Ksenia}\ \bibnamefont {Pougach}}, \ and\
  \bibinfo {author} {\bibfnamefont {Kevin~J}\ \bibnamefont {Verstrepen}},\
  }\bibfield  {title} {\enquote {\bibinfo {title} {Noise and epigenetic
  inheritance of single-cell division times influence population fitness},}\
  }\href@noop {} {\bibfield  {journal} {\bibinfo  {journal} {Current Biology}\
  }\textbf {\bibinfo {volume} {26}},\ \bibinfo {pages} {1138--1147} (\bibinfo
  {year} {2016})}\BibitemShut {NoStop}%
\bibitem [{\citenamefont {Gavagnin}\ \emph {et~al.}(2018)\citenamefont
  {Gavagnin}, \citenamefont {Ford}, \citenamefont {Mort}, \citenamefont
  {Rogers},\ and\ \citenamefont {Yates}}]{gavagnin2018invasion}%
  \BibitemOpen
  \bibfield  {author} {\bibinfo {author} {\bibfnamefont {Enrico}\ \bibnamefont
  {Gavagnin}}, \bibinfo {author} {\bibfnamefont {Matthew~J}\ \bibnamefont
  {Ford}}, \bibinfo {author} {\bibfnamefont {Richard~L}\ \bibnamefont {Mort}},
  \bibinfo {author} {\bibfnamefont {Tim}\ \bibnamefont {Rogers}}, \ and\
  \bibinfo {author} {\bibfnamefont {Christian~A}\ \bibnamefont {Yates}},\
  }\bibfield  {title} {\enquote {\bibinfo {title} {The invasion speed of cell
  migration models with realistic cell cycle time distributions},}\ }\href@noop
  {} {\bibfield  {journal} {\bibinfo  {journal} {arXiv preprint
  arXiv:1806.03140}\ } (\bibinfo {year} {2018})}\BibitemShut {NoStop}%
\end{thebibliography}%
\onecolumngrid
\newcounter{figcounter}
\setcounter{figcounter}{0}
\newenvironment{smfigure}{%
\addtocounter{figure}{-1}
\refstepcounter{figcounter}
\renewcommand\thefigure{S\thefigcounter}
\begin{figure}}
{\end{figure}}

\newcounter{defcounter}
\setcounter{defcounter}{0}
\newenvironment{smequation}{%
\addtocounter{equation}{-1}
\refstepcounter{defcounter}
\renewcommand\theequation{S\thedefcounter}
\begin{equation}}
{\end{equation}}

\section{Supplemental Material}
\subsection{Division time distribution}
Let us start with a new-born cell of size $v_0=\Delta$ (average sized cell) at time $t=0$. Our goal to find the probability density $g_n(v_b, \delta t)$ of an $n$th generation daughter cell being born with the size $v_n = v_b$ at time $t_n = n\bar\tau + \delta t$, where $\bar \tau \equiv \ln(2)/\kappa$ is the doubling time. Using $v_n = v_{n-1} e^{\kappa \tau_n}/2$ and $t_n = t_{n-1}+\tau_n$, we can derive a recursive relationship for $g$. Here $\tau_n$ is given by 
\begin{smequation}
	\tau_n = \frac1\kappa \ln \left(\frac{v_d(v_{n-1})}{v_{n-1}}\right)+\xi_n = \frac1\kappa \ln \left(\frac{2\Delta^\alpha}{v_{n-1}^\alpha}\right)+\xi_n
\end{smequation}%
with the probability density
\begin{smequation}
	f_\tau(\tau|v_b) = f_\xi\left(\tau- \frac1\kappa \ln \left(\frac{2\Delta^\alpha}{v_b^\alpha}\right)\right).
\end{smequation}%
For an $n$th generation cell to be born at time $t_n = n\bar\tau + \delta t$ with size $v_n = v_b$ given the $n$th generation time $\tau_n = \tau$, the $(n-1)$st generation mother cell must have been born with the size $v_{n-1} = 2 v_b e^{-\kappa \tau}$ at time $t_{n-1} = n\bar\tau + \delta t - \tau = (n-1) \bar\tau + (\delta t + \bar \tau - \tau)$, i.e.
\begin{smequation}
\begin{split}
	g_n(v_b, \delta t) &= \int\int g_{n-1}(v_{n-1}, \delta t+\bar \tau - \tau) f_\tau(\tau|v_{n-1}) \delta\left(v_b-v_{n-1} e^{\kappa \tau}/2\right) d v_{n-1}\, d\tau\\
		&= \int 2e^{-\kappa \tau}g_{n-1}(2 v_b e^{-\kappa \tau}, \delta t+\bar \tau - \tau) f_\xi\bigg((1-\alpha)(\underbrace{\tau-\bar \tau}_{\delta \tau})+\frac{\alpha}{\kappa}\ln\left(\frac{v_b}{\Delta}\right)\bigg) d\tau\\
		&= \int e^{-\kappa \delta \tau}g_{n-1}(v_b e^{-\kappa \delta\tau}, \delta t-\delta \tau) f_\xi\left((1-\alpha)\delta \tau+\frac{\alpha}{\kappa}\ln\left(\frac{v_b}{\Delta}\right)\right) d\delta\tau.
\end{split}
\end{smequation}%

Next, we show that $g_n$ has the form $g_n(v_b, \delta t) = \delta\left(v_b - v_0 e^{\kappa\delta t}\right) \tilde g_n(\delta t)$. The initial condition $g_0(v_b, \delta t) = \delta(\delta t) \delta(v_b-v_0)$ satisfies this form. By induction, it is enough to show that $g_{n-1}$ satisfying this form implies that $g_n$ also satisfies this form:
\begin{smequation}
\begin{split}
	g_n(v_b, \delta t) &= \int e^{-\kappa \delta \tau}g_{n-1}(v_b e^{-\kappa \delta\tau}, \delta t-\delta\tau)f_\xi\left((1-\alpha)\delta \tau+\frac{\alpha}{\kappa}\ln\left(\frac{v_b}{\Delta}\right)\right) d\delta\tau\\
		&= \int e^{-\kappa \delta \tau}\delta\left(v_b e^{-\kappa \delta\tau} - v_0 e^{\kappa(\delta t-\delta \tau)}\right) \tilde g_{n-1}(\delta t- \delta \tau) f_\xi\left((1-\alpha)\delta\tau+\frac{\alpha}{\kappa}\ln\left(\frac{v_b}{\Delta}\right)\right) d\delta\tau\\
		&= \delta\left(v_b  - v_0 e^{\kappa\delta t}\right) \underbrace{\int \tilde g_{n-1}(\delta t-\delta\tau) f_\xi\big((1-\alpha)\delta\tau+\alpha \delta t\big) d\delta\tau}_{\tilde g_n(\delta t)}.
\end{split}
\end{smequation}%

Now, we have the recursive relationship (dropping the tildes)
\begin{smequation}
	g_n(\delta t) = \int g_{n-1}(\delta t-\delta\tau) f_\xi\big((1-\alpha)\delta\tau+\alpha \delta t\big) d\delta\tau
\end{smequation}%
for the distributions $g_n$ of the deviation $\delta t$ from the expected timing of $n$th division. This relationship is derived starting with a new-born cell of size $\Delta$ at time $t=0$. A cell larger (smaller) than $\Delta$ is expected to have an $n$th division time earlier (later) than $n \bar \tau$. Redefining $\delta t$ such that $t_n = n\bar \tau - \alpha\ln(v_0/\Delta)/\kappa + \delta t$ would yield the same recursive relationship for the distribution of $\delta t$ starting with an arbitrary initial size $v_0$.\\

\vspace{-1em}
\subsection{Effect of noise in the division ratio on the decay rate of the oscillations}
Let us define the division ratio $r$ to be the ratio of the size of a daughter cell to that of its mother cell right before division. For a symmetrically dividing organism, it has the average $\bar r = 0.5$. The variance $\sigma_r^2$ of $r$ is typically much smaller than that of the single-cell growth rate, and therefore, it was neglected in the discussion in the main text. However, this variance could vary for different organisms and different growth conditions. To compute the effect of random variations in the division ratio on the decay rate of the oscillations, we take advantage of the main conclusion of the paper, that is the variables $\alpha$ and $\sigma_\xi$ do not play a role on the relaxation time of the population. Using these results, we can evaluate the effect of noise in division ratio on the relaxation rate for the simplest case of $\alpha =1$ and $\sigma_\xi=0$ and compare the predictions with simulations with realistic values of $\alpha$ and $\sigma_\xi$.\\

For the simple case of $\alpha =1$ and $\sigma_\xi=0$, each cell divides precisely when its size reaches the finial size of $2\Delta$. The size of the new born cell is given by $2r\Delta$. If $r$ is different from $\bar r = 1/2$ due to noise in division, the new born cell will be slightly different in size compared to the average cell and divide with slightly different generation time $\tau$. However, this has the exact same effect on the desynchronization of the population as if the cell would divide symmetrically but had the same delay in its next generation due to a small noise in its growth rate $\kappa$ instead. We already know how much noise in $\kappa$ affects the relaxation rate, so if we compute how much of noise $\delta r$ in $r$ gives rise to the same amount change $\delta \tau$ that is due to some noise $\delta \kappa$ in $\kappa$, then we know how the variability in $r$ affects the relaxation rate $\lambda$.\\

The generation time from Eq. (1) of the main text for this case simplifies to $\tau = -\ln(r)/\kappa$. The relative change in the generation time, $\delta \tau/\bar\tau$, due to a small deviation $\delta \kappa$ in $\kappa$ is given by $\delta\tau/\tau = -\delta \kappa/\bar\kappa$. Similarly, the relative change in the generation time due to the a small change $\delta r$ in $r$ is given by $\delta\tau/\tau = \delta r/(\bar r\ln(\bar r))$. In the absence of variability in division ratio, the relaxation time is given by $\lambda/\bar\kappa = C\left(\sigma_\kappa^2/\bar\kappa^2 \right)$ (see Fig.~4 of the main text). The above argument suggest that the relaxation time with the stochastic division ratio should be given by
\begin{smequation}
	\frac{\lambda}{\bar\kappa} = C\left(\frac{\sigma_\kappa^2}{\bar\kappa^2}+\frac{\sigma_r^2}{(\bar r\ln(\bar r))^2} \right)
		= C\left(\frac{\sigma_\kappa^2}{\bar\kappa^2}+\frac{4\sigma_r^2}{\ln^2(2)} \right).
		\label{eq:division_ratio}
\end{smequation}%
The constant $C$ was measured in simulations in the main text and has the value $C\approx 29$. \Fig{division_ratio} compares this prediction with simulation results with realistic values of $\alpha$ and $\sigma_\xi$ showing that this relationship indeed holds independent of $\alpha$ and $\sigma_\xi$.
\begin{smfigure}[!h]
	\centering
	\includegraphics[width=0.55\columnwidth]{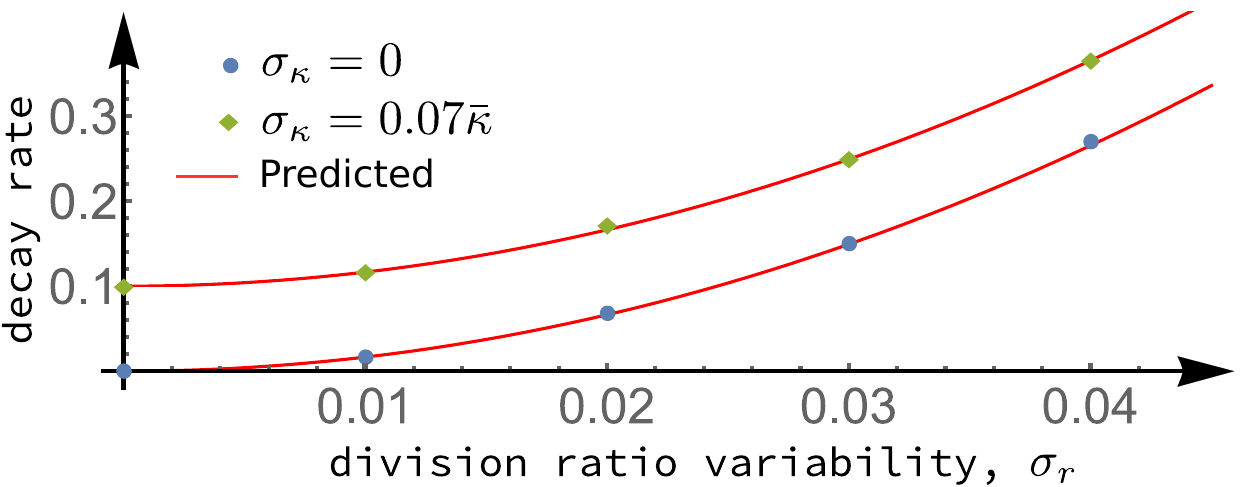}
	\caption{Simulation results (markers) for the effect of variability in the division ratio on the decay rate of the oscillations compared with prediction from \eq{division_ratio} (no fitting parameter, the value of $C$ is taken from the fit to $\sigma_r = 0$ case from Fig.~4 of the main text). Simulation parameters: $\bar \tau = 1$, $\alpha=0.5$, $\sigma_\xi = 0.1 \bar\tau$, $\sigma_\kappa \in \{0, 0.07\bar\kappa\}$, and $\sigma_r \in \{0, 0.01,0.02,0.03, 0.04\}$.}
	\label{fig:division_ratio}
\end{smfigure}

\subsection{Relationship between the steady-state population growth rate and single-cell growth rate distribution}
There is a well-known equation by Powell~\cite{powell1956growth} for the relationship between the growth rate of the population and the distribution of the generation times, assuming that the generation times are not correlated. This equations reads:
\begin{smequation}
	\int_0^\infty f_\tau(\tau) e^{-k \tau} d\tau= \frac12,
\end{smequation}%
where $f_\tau$ is the probability density function of the generation times $\tau$. Similar to the previous section, we can use the main results of the paper that $\alpha$ and $\sigma_\xi$ do not play a role in the dynamics of the population growth rate to simplify the problem. We can set $\alpha = 1$ and $\sigma_\xi = 0$ and find the relationship between the population growth rate and the distribution of single-cell growth rates and confirm with simulations that it indeed holds for other value of $\alpha$ and $\sigma_\xi$. At this limit, the generation time $\tau$ is given by $\ln(2)/\kappa$ and the generation times of mother and daughter cells are uncorrelated, and therefore, Powell's relationship applies. Moreover, we have $f_\tau(\tau) d\tau =  \rho(\kappa) d\kappa$. Therefore, Powell's relationship simplifies to 
\begin{smequation}\label{eq:rho}
	\int_0^\infty \rho(\kappa) e^{-\ln(2)\frac{k}{\kappa}} d\kappa
		= \int_0^\infty \rho(\kappa) 2^{-\frac{k}{\kappa}} d\kappa
		\equiv \mean{\left(\frac12\right)^{k/\kappa}}_\kappa = \frac12.
\end{smequation}%
We test the prediction of this relationship for two classes of growth rate distributions, gamma distribution and inverse gamma distribution. For both of these distribution, it can be shown through a scaling argument of \eq{rho} that the ratio $k/\bar \kappa$ is only a function of $\sigma_\kappa/\bar\kappa$. \Eq{rho} can be numerically solved to find these relationships. \Fig{steady} compares this functional dependence predicted from \eq{rho} to its values obtained by simulation. Both parameters $\alpha$ and $\sigma_\xi$ are varied along with $\sigma_\kappa$ to show that they do not play a role in the value of the population growth rate $k$.
\begin{smfigure}[!h]
	\centering
	\includegraphics[width=0.55\columnwidth]{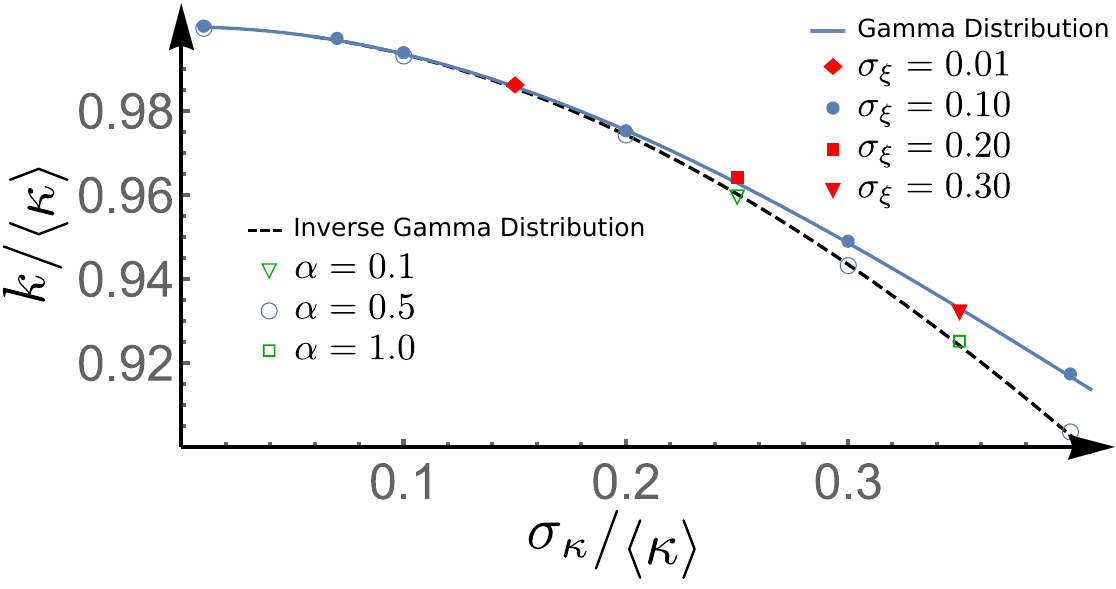}
	\caption{Predicted dependence of the population growth rate $k$ on the distribution of single-cell growth rates $\kappa$ for two classes of distributions, gamma distribution (solid blue line) and inverse gamma distribution (dashed black line) compared with simulation results (markers). The values of $\sigma_\xi$ and $\alpha$ are changed along with $\sigma_\kappa$ for the gamma and inverse gamma distributions, respectively, to show that they do not affect the of population growth rate.}
	\label{fig:steady}
\end{smfigure}

\vspace{-2em}
\subsection{Experimental verification on single-cell data from the \lq\lq mother machine" experiment~\cite{wang2010robust}}
Although we are not aware of any experimental data directly measuring the dynamics of the growth rate of a population starting from a single cell, we can confirm the predictions of our model on an ensemble of single-cell lineages as are measured in the \lq\lq mother machine" experiment~\cite{wang2010robust} (data courtesy of Suckjoon Jun's group). We use the data on the growth and divisions of long lineages of the wild-type strains of \textit{Escherichia coli} MG1655 only keeping track of the cells inheriting the old pole (see Ref.~\cite{wang2010robust} for details). The data set includes $113$ cells with at least $36$ generations. If we synchronize the first division of each cell and histogram the division events over time, the division times of cells desynchronize after a couple of generations (see the left panel of \fig{shift}). This is not surprising; the cells are not descendants of the same cell, and in particular, the first generation cells do not have the same size. However, as discussed in the last paragraph of Section A, if a cell is born with a size $v_0$ different from the average size $\Delta$, the expected value of its division times at long time get shifted by $-\alpha \ln(v_0/\Delta)/\kappa$. To test this prediction, we can shift back the time for each lineage by $+\alpha \ln(v_0/\Delta)/\kappa$ based on the initial size of its first generation and recompute the histograms. As shown in the right panel of \fig{shift}, after this shift, the division times of the cells stay synchronized for a long time. 

\begin{smfigure}[t]
	\centering
	\hspace{-2em}
	\includegraphics[width=0.49\columnwidth]{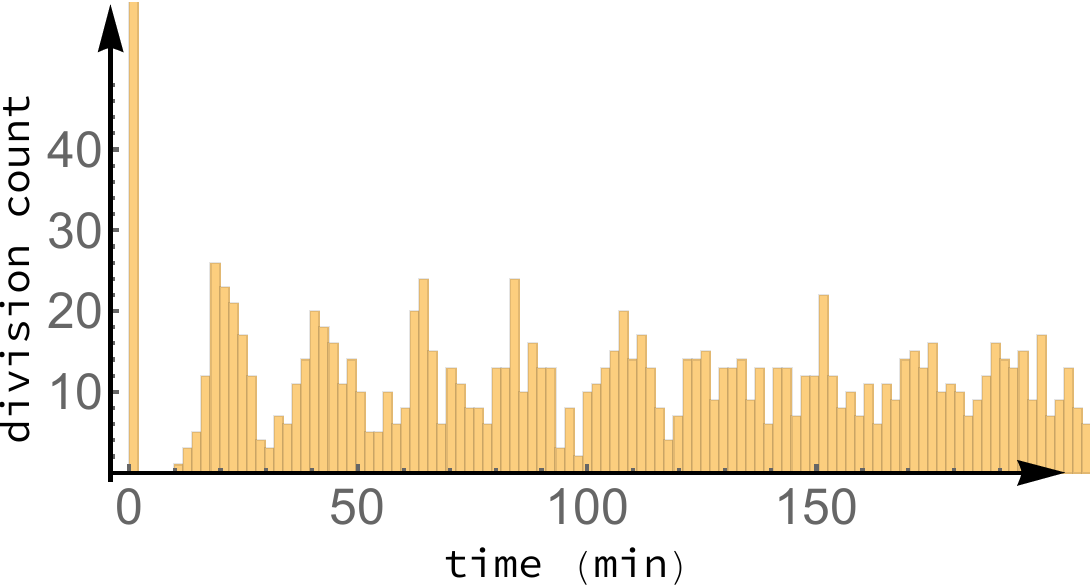}
	\includegraphics[width=0.49\columnwidth]{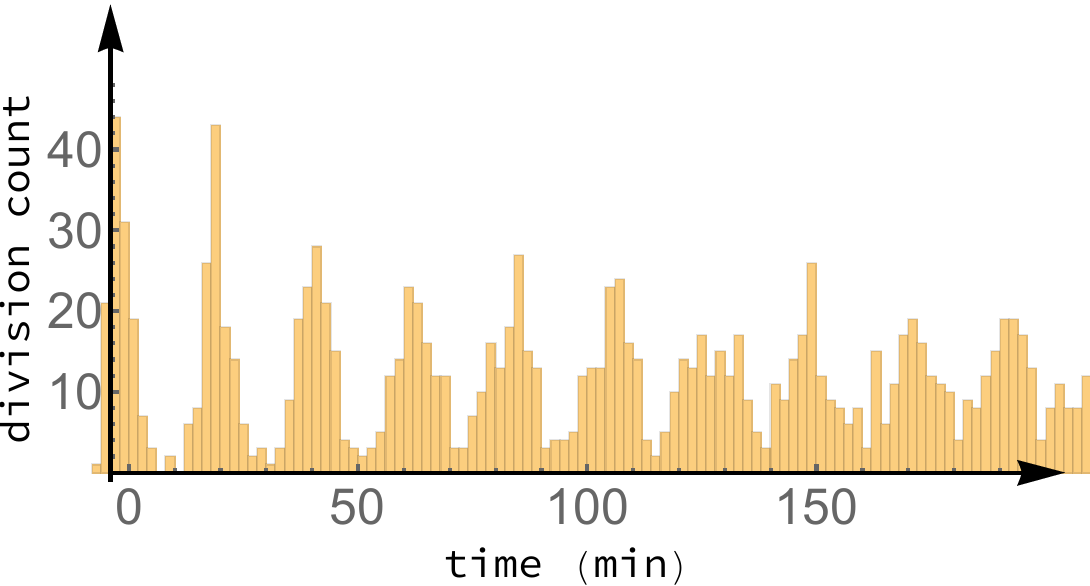}
	\caption{Histogram of the timing of successive divisions for an ensemble of cells. After each division, only the cell inheriting the old pole is kept in the population. (Left) The time for each lineage (first generation cell and all of its progenies) was shifted so that the first division for all the lineages take place at time zero. Different lineages become desynchronized after a few generations due to the difference in initial cell sizes. (Right) The time is shifted for each lineage by $+\alpha \ln(v_0/\Delta)/\kappa$ based on the initial size of the first generation cell $v_0$ to recover the sustained oscillations analogous to the oscillation in the growth rate of a population starting from a single cell. The data is taken from Ref.~\cite{wang2010robust}.}\vspace{-1em}
	\label{fig:shift}
\end{smfigure}

The rate at which the division times of different cells become desynchronized should determine the variance of the single-cell growth rate distribution. We find this rate by fitting an exponentially decaying sine curve to the histogram of the division times at long time as shown in \fig{fit}. The measured decay rate was found to be $\lambda\approx (0.22 \pm 0.04)~ hour^{-1}$. The mean of single-cell growth rates was measured to be $\bar \kappa=(1.840\pm0.001)~hour^{-1}$. The predicted standard deviation of single-cell growth rates is found to be 
\begin{smequation}
	\sigma_{\kappa, predicted}=\sqrt{\bar\kappa \lambda/C}=(0.12\pm0.01)~hour^{-1},
	\label{eq:prediction}
\end{smequation}%
which is in an excellent agreement with the direct measurement of sample standard deviation, $\sigma_{\kappa, measured}=0.135~hour^{-1}$. The value of $C$ in \eq{prediction} is taken from the simulation fit in the main text, $C\approx 29$. The measured and predicted coefficients of variation of single-cell growth rates are $CV_{measured}=0.073$ and $CV_{predicted}=0.064\pm0.006$. Single-cell growth rates are measured by linear fits to the logarithm of cell lengths as functions of time. Fits with $R^2$ less than $0.98$ were excluded.

In the absence of the feedback from cell-size control, the oscillations would decay much more rapidly, and $\sigma_{\kappa, predicted}$ calculated from the decay rate using \eq{prediction} would be much larger than the sample standard deviation of single-cell growth rates. However, in our case, $\sigma_{\kappa, predicted}$ is slightly smaller than the direct sample measurement. Note that the sample standard deviation is biased due to measurement error and is always slightly higher than the actual standard deviation (for large samples), and therefore, we believe the value predicted from the decay rate is more accurate than the sample standard deviation. 
\begin{smfigure}[!h]
	\centering
	\vspace{-1em}
	\hspace{-2em}
	\includegraphics[width=\columnwidth]{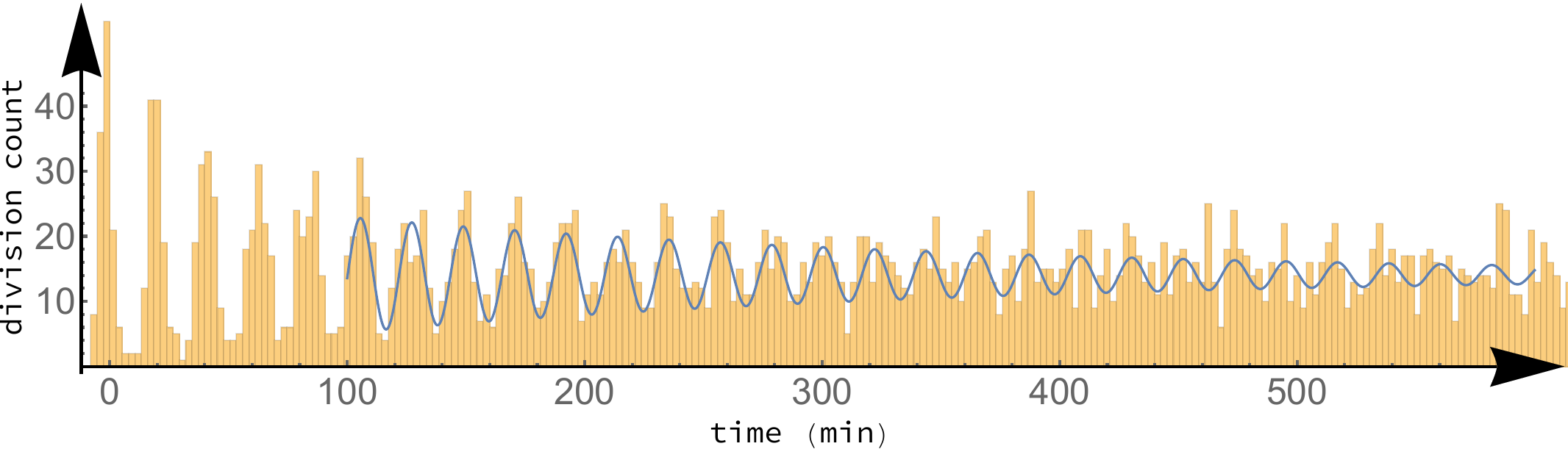}
	\caption{Decaying sinusoidal fit to the long-time dynamics of division rates of the ensemble discussed in \fig{shift}. The decay rate of these oscillations is measure to be $\lambda\approx (0.22 \pm 0.04)~ hour^{-1}$ which determines the variability in single-cell growth rates of $\sigma_{\kappa}=(0.12\pm0.01)~hour^{-1}$ compared to direct measurement of the sample standard deviation $\sigma_{\kappa}=0.135~hour^{-1}$.}\vspace{-3em}
	\label{fig:fit}
\end{smfigure}

\end{document}